\documentclass[aps,twocolumn,eqsecnum,amsmath,amssymb,floatfix]{revtex4}
\usepackage{epsfig,psfrag}
\begin{document}
%\draft
%%%%%%%%%%%%%%%%%%%%%%%%%%%%%%%%%%%%%%%%%%%%%%%%%%%%%%%
\title{Inter edge Tunneling in Quantum Hall Line Junctions}
%%%%%%%%%%%%%%%%%%%%%%%%%%%%%%%%%%%%%%%%%%%%%%%%%%%%%%%
\author{Eun-Ah Kim}
\author{Eduardo Fradkin}

\affiliation{Department of Physics, University of Illinois at
Urbana-Champaign, 1110
W.\ Green St.\ , Urbana, IL  61801-3080, USA}
\date{\today}

%%%% ALIASES and NEWCOMMAND%%%%%%%%%%%%%%%%%%%%%%%%%%%%%%%%%%%%%%%%%%
\newcommand{\lang} {\langle}
\newcommand{\rang} {\rangle}
\newcommand{\tphi} {{\tilde{\phi}}}
\newcommand{\ttheta} {{\tilde{\theta}}}
\newcommand{\tPi} {{\tilde{\Pi}}}
\newcommand{\tphir} {{\tilde{\phi}}_+}
\newcommand{\tphil} {{\tilde{\phi}}_-}
\newcommand{\D}{\displaystyle}
%%%%%%%%%%%%%%%%%%%%%%%%%%%%%%%%%%%%%%%%%%%%%%%%%%%%%%%%%%%%%%%%%%%%%%%

\begin{abstract}
We propose a scenario to understand the puzzling features of the
recent experiment by Kang and coworkers on tunneling between laterally coupled quantum Hall liquids
by modeling the system as a pair of coupled chiral Luttinger liquid with a point
contact tunneling center. We show that for filling factors $\nu\!\sim\!1$ the 
effects of the Coulomb interactions move the system deep into strong tunneling regime,
by reducing the magnitude of the Luttinger parameter $K$, leading to the appearance 
of a zero-bias differential conductance peak of magnitude $G_t\!=\!Ke^2/h$ at zero temperature. 
The abrupt appearance of the zero bias peak as the filling factor is increased past a value 
$ \nu^*\! \gtrsim 1$, and its gradual disappearance thereafter can be understood as a 
crossover controlled by the main
energy scales of this system: the bias voltage $V$, the crossover scale $T_K$, 
and the temperature $T$. The low height of the zero bias peak
$\sim 0.1e^2/h$ observed in the experiment, and its broad finite width, 
can be understood naturally within this picture. Also, the abrupt reappearance of the zero-bias 
peak  for $\nu \gtrsim 2$ can be explained as an effect caused by spin reversed electrons, \textit{
i.\ e.\/} if the 2DEG is assumed to have a small polarization near 
$\nu\!\sim\!2$.  
We also predict that as the temperature is lowered $\nu^*$ should decrease, and the width of
zero-bias peak should become wider. This picture also predicts the existence of 
similar zero bias peak in the spin tunneling conductance near for $\nu \gtrsim 2$.
\end{abstract}
\maketitle
%%%%%%%%%%%%%%%%%%%%%%%%%%%%%%%%%%%%%%%%%%%%%%%%%%%%%%%%%%%%%%%%%%%%%%%

%\pacs{PACS numbers: 71.45.Lr, 72.15.-v, 73.23.-b}

% 71.45.Lr Charge-density-wave systems
% 72.10.Bg General formulation of transport theory
% 72.15.Nj Collective modes
% 72.15.-v Electronic conduction in metals and alloys
% 73.23.-b Mesoscopic systems
% 73.23.Ad Ballistic transport
% 73.20.Dx Electronic states in low-dimensional structures
% 73.40.Rw Metal-insulator-metal structures
% 05.60.-k Transport processes
% 11.30.Rd Chiral symmetries
% 11.40.Dw General theory of currents

%\begin{multicols}{2}

%\narrowtext

%%%%%%%%%%%%%%%%%%%%%%%%%%%%%%%%%%%%%%%%%%%%%%%%%%%%%%%%%%%%%%%%%%%%%%%
%\section{Introduction}
%%%%%%%%%%%%%%%%%%%%%%%%%%%%%%%%%%%%%%%%%%%%%%%%%%%%%%%%%%%%%%%%%%%%%%%

The properties of the edge states of two-dimensional electron gases (2DEGs) 
in high magnetic fields reflect the structure of the Hilbert spaces of bulk 
fractional, and integer, quantum Hall (FQH) states. 
In the absence of edge reconstruction, the low energy Hilbert spaces of
FQH edge states can be represented by a suitable set of 
chiral Luttinger liquids~\cite{hal,wen,stone}. 
This identification brought considerable interest in the study of FQH edge 
states as a well controlled laboratory for experimental exploration of quantum 
transport properties of Luttinger liquids. Much effort has been devoted to the 
theoretical~\cite{kane,extra-kane} and experimental study of tunneling of both
between FQH edge states~\cite{milliken} and into FQH edge states~\cite{chang}. 
Measurements~\cite{chang} of electron tunneling from a bulk doped-GaAs
electron into the sharp edge of a FQH state with filling fractions $\nu \leq 1$ 
have confirmed the existence of both the scaling regime~\cite{kane,extra-kane} 
and the crossover behavior~\cite{chamon} predicted by the chiral Luttinger liquid picture. 
However, many important open questions remain about the actual observed behavior of the 
tunneling exponent and its consistency with the physics of the bulk FQH states (see, for 
instance 
~\cite{kane,shytov,ana,joel}, and references therein).

Recently, W.\ Kang and coworkers ~\cite{kang} have measured the differential tunneling 
conductance 
of a device in which two 2DEG's in the integer quantum Hall regime are laterally 
coupled through an atomically precise tunneling barrier.  
Their data shows a very sharp and intense differential conductance peak of height 
$G_t\equiv dI_t/dV\approx 0.1e^2/h$ at zero bias for certain ranges of magnetic field on 
top of an oscillatory behavior, which appears in qualitatively same manner for all range 
of magnetic field. The data shows an abrupt appearance and the following gradual 
disappearance of the zero bias conductance (ZBC) peak as the filling factor is increased past apparent 
threshold values $\nu^*_1 \gtrsim 1$ and $\nu^*_2 \gtrsim 2$ respectively. 
In both cases, the height of the ZBC peak they observed is considerably 
smaller than the quantum of conductance $e^2/h$ and the ZBC 
peak was observed over a fairly broad range of filling fractions ($\lesssim e^2/2h$). 
The data of Kang {\it et.\ al.\/}~\cite{kang} shows no ZBC peak in the tunneling 
conductance for $\nu\leq 1$.

The theoretical explanation of the experiment of Kang and coworkers has focused on the 
fact that 
it is not
possible to tunnel electrons between two perfectly aligned FQH edges with opposite 
chirality~\cite{wen}.
Thus, if the barrier is assumed to be atomically precise, the only way in which 
tunneling can possibly
take place is by the anti-crossing of Landau levels belonging to both sides of the 
barrier~\cite{ho}. 
In the Landau gauge ${\vec A}\!=\!(0,Bx,0)$, where $x$ direction is chosen perpendicular 
to the
barrier and $y$ direction along the barrier, the single particle
wave function has a form $\varphi(x,y)\!=\!\exp(ik y)\phi_k(x)$ where
$\phi_k(x)$ is an eigenfunction of the Hamiltonian
$H_k(x)\!=\!-\frac{\hbar^2}{2m}\frac{\partial^2}{\partial
x^2}\!+\!\frac{1}{2}m\omega_c^2(x\!-\!kl^2)^2\!+\!V_B(x)$, with $V_B(x)$ a
potential due to barrier which is symmetric about $x\!=\!0$. The
dispersion curves originating from the two systems on both sides
of the barrier overlap around $k\!=\!0$. At the crossing points,
gaps open as a consequence of a coupling between the counter
propagating edge states~\cite{ho}. 
This is indeed
the scenario assumed in the work of Kang and coworkers~\cite{kang} and by 
Mitra and Girvin~\cite{girvin},
Lee and Yang~\cite{lee-yang}, 
Kollar and Sachdev~\cite{sachdev}, and by an earlier calculation by 
Takagaki and Ploog~\cite{takagaki}.

In this picture, the appearance of a zero-bias conductance peak is ascribed to the 
existence of a gap 
in the spectrum of edge states at the barrier, since a gap suppresses the conduction 
channel along the barrier 
provided by unmixed edge states with opposite chirality formed by the barrier. 
Mitra and Girvin~\cite{girvin}, as well as Kollar and
Sachdev~\cite{sachdev}, observed that electron-electron interactions yield a 
substantial modification of
the gap which cannot be accounted for by level mixing arguments.
In these theories, the gap is equal to the soliton energy of a quantum sine-Gordon model,
derived from a microscopic theory of the barrier. Notice that, due to the Landau level 
mixing induced by the
barrier, the effective Fermi wave vector of the barrier states is $k_F=0$. 
Thus a gap in the spectrum does
not require backscattering in this geometry.
In particular, Mitra and Girvin~\cite{girvin} used a Hartree-Fock
theory to calculate the Luttinger liquid parameter, the collective
mode velocity, and the momentum cutoff of the effective sine-Gordon theory. 
It was found that Coulomb
interaction, which is taken into account in Hartree-Fock,
leads to a substantial enhancement of the gap. More recently, Kollar and
Sachdev~\cite{sachdev}, used a method of matched asymptotics 
to determine the momentum cutoff for sine-Gordon theory. The
gap they found is larger than the result of Mitra and Girvin. 

However, even with the gap obtained by Kollar and Sachdev ~\cite{sachdev} it is not 
possible to understand the height of the zero-bias conductance peak. 
Both references ~\cite{girvin} and ~\cite{sachdev} predict on
general grounds a zero-bias peak with height $e^2/h$, larger than the experimental 
result $0.1 e^2/h$ of Ref.~\cite{kang} by approximately one order of magnitude.
Furthermore, in this picture the ZBC peak is expected
above the second Landau level in the non-interacting system,
whereas peak region was prominent near $\nu^*\approx1$ in the
experiment. (Interaction effects do not modified this result in any essential way.) 
Given these facts it
was argued in Refs.\ ~\cite{girvin} and ~\cite{sachdev} that effects of disorder may 
be ultimately responsible for these discrepancies between theory and experiment.

In search of an answer to these questions, we reexamined the alternative scenario of tunneling 
between counter-circulating edge states through an imperfection of the tunneling barrier. 
We were motivated partly by the observation that the effects of anti-crossing induced 
by the barrier are not expected to occur at least before the second Landau level begins to be 
filled, which is not the regime in which the zero-bias
peak first appears. Thus we will assume the more standard situation of a barrier separating two
FQH states with edges of opposite chirality and non-vanishing Fermi wave-vectors. 
Under these circumstances tunneling is only allowed if impurities and imperfections are present. 
This is a possibility that must be considered seriously particularly given that in the end 
impurity scattering is invoked as the explanation for the
magnitude of the zero-bias peak, as advocated in Refs.\ ~\cite{girvin} and ~\cite{sachdev}. 
Thus, in this
paper we will assume that the barrier is precise enough to have just a few imperfections
which act as weak tunneling centers. In fact we will assume that there is just one such 
tunneling center.

In the situation of the experiment of 
Kang {\it et.\ al.\/}, where right and left moving edges
were spatially separated by a barrier, a local deformation of the 
edges due for instance to an impurity can
result in a weak tunneling center which mimics the pinch-off
effect of the patterned back gate electrode of the
experiment by Milliken, Umbach and Webb~\cite{milliken}. The
authors of Ref.~\cite{milliken} have observed expected temperature
dependence of tunneling conductance through point
contact~\cite{moon,wen9091} for $\nu=1/3$. However, a quite
unique feature of the set up of Ref.\ ~\cite{kang} is that it can explore not
only the effect of back scattering through a (presumably) point
contact,  but also the effects of electron-electron interactions along the edges. 
 
Our analysis shows that the electron-electron interaction
plays a crucial role in the tunneling conductance. Electron-electron interactions 
turn the pair of edge states into a single non-chiral Luttinger liquid with an effective 
Luttinger parameter $K <1$ for filling factors $\nu \gtrsim 1$. This problem can be 
mapped into the problem of a junction in a Luttinger liquid first 
studied by Kane and Fisher~\cite{kane,extra-kane}, with a Luttinger parameter
reduced from $1$ due to the effects of the Coulomb interactions along the barrier,
which brings the system to strong tunneling phase if it were at $T\!=\!0$. 
In references ~\cite{kane} and
~\cite{extra-kane}, Kane and Fisher pointed out that for $K < 1$, tunneling  
at a point contact is a
relevant perturbation and the system flows to a strong coupling regime. 
While $K<1$ suggests that the threshold for a zero-bias peak should be 
observed at a filling factor somewhat below $\nu=1$, we find that there 
is a non-trivial temperature dependence of the height and width of the 
zero-bias peak induced by the renormalization flow of the tunneling operator.

We studied the effects of finite temperature by 
mapping the problem to boundary sine-Gordon (BSG)
problem which is exactly solvable. By combining a number of known exact 
results of the BSG theory with the calculation of an 
appropriate renormalization group $\beta$-function, we suggest a natural 
explanation of the salient features of the experiment of Ref.\ ~\cite{kang}. 
We studied in detail the crossover behavior of the tunneling conductance as a 
function of temperature and found that it can explain qualitatively the observations of 
Ref.\ ~\cite{kang}. 
We find that  finite temperature is responsible for both the low height of 
the peak and its gradual disappearance when the filling factor is increased 
past $\nu\!\sim\!1$. Further experimental studies of the temperature 
dependence of the zero-bias peak can check these theoretical predictions. 
In particular we give an explicit expression for the temperature dependence of 
the differential conductance at zero bias voltage for the particular value of the 
Luttinger parameter $K=1/2$. For more general values of the Luttinger parameter the solutions are
more complicated but nevertheless vary smoothly and slowly with $K$ (see below).
Although the data that has been published so far of the experiment of Kang and 
coworkers~\cite{kang} is at a temperature of $300 \; mK$, unpublished data from the same group 
in the temperature range $300 \;
mK$ to $8 \; K$ is well described by our results~\cite{kang-priv}.  

We have also studied tunnel junctions at a barrier in  partially 
spin-polarized QH states. We find that the reappearance of the peak region near $\nu\!\sim\!2$ can be  
explained if the electron gas is not fully polarized but instead has a small 
spin polarization. We also consider in this paper the interesting case of a line junction in a 
spin singlet $\nu=2$ state. We find that for these QH states, at 
$\nu \gtrsim 2$ a spin-spin interaction across a single point junction leads to a number of
interesting effects in both spin and charge transport across the junction.

This paper is organized as follows. In Sec.~\ref{sec:model}, we introduce the 
model for a IQH-barrier-IQH junction with a single tunneling center and bosonize the 
model. In Section \ref{sec:BSG} we map the model to the integrable BSG model by using a 
standard  folding 
procedure. The result will be used to understand the experiment near $\nu\!=\!1$ . 
In Section \ref{sec:spinful} we 
propose an explanation for the experimental results near $\nu\!=\!2$
with the assumption that there is a small spin polarization for $\nu\!=\!2$. Here we generalize our analysis and discuss the role of exchange, Zeeman and magnetic anisotropy interactions on tunneling processes.
Finally, in  Section \ref{sec:conclusion} we review our main results and give 
some predictions on future experiment based on our analysis.

%%%%%%%%%%%%%%%%%%%%%%%%%%%%%%%%%%%%%%%%%%%%%%%%%%%%%%%%%%%%%%%%%%%%%%%%%%%%%%%%%%%%%%%
\section{The model Hamiltonian}
\label{sec:model}
%%%%%%%%%%%%%%%%%%%%%%%%%%%%%%%%%%%%%%%%%%%%%%%%%%%%%%%%%%%%%%%%%%%%%%%%%%%%%%%%%%%%%%

We begin by briefly describing the experimental setup and the most salient 
results of ref.\ ~\cite{kang}. The 2DEG-barrier-2DEG junctions used by Kang 
{\it et.\ al.\/} ~\cite{kang} consisted of two regions of 2DEG of widths
$13\mu$\@m and $14\mu$\@m, where the electrons live in the
two-dimensional interface of the GaAs-AlGaAs heterostructure,
separated by 88\AA- thick $\text{Al}_{0.1}\text{Ga}_{0.9}\text{As}
$/AlAs barrier of height 220 meV. These junctions are believed to be atomically precise, 
which means that they have very few defects on their entire length. 
In the experiment the conductance at $T=300\text{mK}$ showed an
oscillatory behavior as a function of bias voltage with successive 
peaks spaced by an
energy of the order of the cyclotron energy $\hbar\omega_c$
in the full range of magnetic field. This effect suggests that there is a  
mixing between Landau levels enabled
by a level shift due to large bias voltage. However, for
fillings $\nu\equiv nh/eB\gtrsim 1$ and $\nu\gtrsim 2$, a sharp
conductance peak dominates at zero bias. The peak heights were
$0.12e^2/h$ and $0.11e^2/h$ respectively for the samples published,
but the height typically varies from sample to sample, always
being of the order of $0.1e^2/h$~\cite{kang-priv}.

%%%%%%%%%%%%%%%%%%%%%%%%%%%%%%%%%%%%%%%%%%%%%%%%%%%%%%%%%%%%%%%%%%%%%%%%%%%%%%%%
\begin{figure}[!b]
\begin{center}
\leavevmode
%\vspace{.2cm}
\noindent
%\hspace{1.0 in}
\epsfxsize=.30\textwidth
\epsfbox{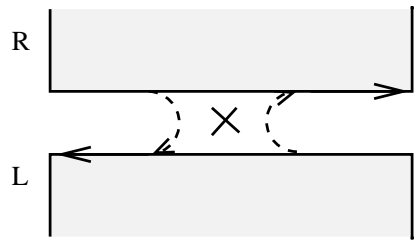}
%\vspace{.5cm}
\end{center}
\caption
{A line junction  with a single backscattering center. 
The two shaded regions and the space
  between correspond respectively to two regions of 2DEG of widths $13\mu$\@m
  and $14\mu$\@ and the 88\AA- thick ${\text Al}_{0.1}{\text Ga}_{0.9}{\text
  As}$/AlAs barrier of 2DEG-barrier-2DEG junctions used by Kang {\it et
  al.}. The  single tunneling center is represented by a cross in the figure. 
  The
  system is equivalent to a one-dimensional Fermi system with right and left 
  moving branches,
  interacting with each other through short range interactions.}
\label{fig:ljunc}
\end{figure}

%%%%%%%%%%%%%%%%%%%%%%%%%%%%%%%%%%%%%%%%%%%%%%%%%%%%%%%%%%%%%%%%%%%%%%%%%%%%%%%%
The model Hamiltonian for the set up of the experiment of Kang and coworkers 
that we will use here is a variant of the one considered by Kane and 
Fisher~\cite{kane,extra-kane}. 
We will make the simplifying assumption the electron-electron 
interactions at the barrier are sufficiently well screened so that they can be represented 
by effective short range intra-edge and inter-edge interactions. While this assumption is 
not fully justified it represents a minor change to the physics of the system. 
Thus, the effects of the width of the barrier are included in the matrix element. 
The right and left moving branches represent the edge states of two $\nu=1$ QH states 
laterally coupled by the 
barrier. These edges have non-vanishing  Fermi wavevectors equal in magnitude (for a 
symmetric barrier) and with opposite direction indicating the chiral nature of the 
edge states. Back scattering is forbidden everywhere 
due to momentum conservation and in the absence of a  periodic potential there is 
no umklapp scattering. The electron-electron interactions are thus purely due to 
``forward scattering" both intra and inter edge, which conserve chirality. 
Thus, under these assumptions, the pair of edge states behaves effectively like a 
single non-chiral one-dimensional Luttinger liquid, with an effective velocity 
$v_0$ and an effective Luttinger coupling constant $g_c$. 
The main effect of the impurity is to provide for a backscattering center at the  
impurity site which we will define to be the origin, $x=0$. The model that we will discuss and solve 
for two coupled $\nu=1$ edges with opposite chirality can be easily extended to discuss the same 
issues for 
fractional quantum Hall states. However, for reasonable values of the dimensionless coupling 
constant (defined below) 
the resulting effective Luttinger parameter is always in the range $K>1$ in which tunneling 
is suppressed and no ZBC 
peak can be observed. Thus, for the rest of this paper we will restrict our discussion to 
the case $\nu>1$ in which there are 
no fractional quantum Hall states (for fully polarized systems).

The system can thus be treated as if it were
effectively one-dimensional, \textit{i.\ e.\/} as if the right and left moving branches
overlapped with each other, and were coupled via a screened Coulomb interaction.
Following Wen's hydrodynamic approach~\cite{wen,wen9091}, the edge states of oppositely moving
modes are described in terms of normal ordered right and left moving densities $J_{\pm}$ 
which satisfy equal-time commutation relations in the form of a $U(1)$ Kac-Moody algebra:
\begin{equation}
[J_{\pm}(x),J_{\pm}(x')]=\mp\frac{i}{2\pi}\partial_x\delta(x-x')
\end{equation}
The Hamiltonian density for the line junction may be written as
a sum of two terms ${\mathcal H}={\mathcal H}_G+{\mathcal H}_t$, where ${\mathcal H}_G$ 
includes the effects of both inter and intra edge 
interactions, and ${\mathcal H}_t$ represents tunneling term at
$x=0$. ${\mathcal H}_G$ is given by 
\begin{equation}\label{HG}
{\mathcal H}_G=\pi v_0(J_-^2+J_+^2+2g_cJ_+J_-),
\end{equation}
where we assumed the speed of right and left
moving electrons to be same with $v_0$, and the third term stands
for the density-density interaction between chiral electrons. 

The dimensionless coupling constant $g_c$, which measures
the strength of the interaction, can be estimated to be $g_c\sim
U/E_F$ where, for the case of Coulomb interactions,
$U\equiv e^2/\epsilon d$ where $d$ is the effective distance between the two edges, 
$\epsilon$ is the static
dielectric constant, and $E_F$ is the Fermi energy for the edge states, assumed to be 
the same on both sides of barrier. 
%For the parameters
%relevant to the experiment~\cite{kang,kang-priv}: $d=88 \AA $,
%$E_F=7.1\text {meV}$ and $\epsilon=12.86$~\cite{sadao}, $U/E_F\sim
%0.896$ which is not small. 
It is important to keep in mind that in practice there is 
no reliable way to determine $g_c$ in terms of microscopic parameters. 
Still, this lowest order estimation implies that Coulomb
interaction must be fairly strong in the actual experimental setup.
In any case, we expect that the dimensionless coupling constant $g_c$ should be a smooth
function of the bulk filling factor $\nu$ and  of the thickness of the barrier.
Intuitively we expect that as the filling factor increases, either by raising the electron
density or by decreasing the magnetic field, the effective distance between the edges of
the two quantum Hall liquids will decrease. Consequently we expect that the
dimensionless coupling constant $g_c$ will increase as the filling factor increases.
We will see below that this effect will play an important role in the explanation of 
the effects seen in the experiments of Kang and coworkers~\cite{kang}.
 
We will represent the effects of back-scattering at the tunneling center (at the origin) 
by a local tunneling operator which in terms of right and left 
moving electron creation and annihilation operators has the standard form 
\begin{equation}
{\mathcal H}_t=t(\psi_+^\dagger \psi_-+\psi_-^\dagger \psi_+)\delta(x)
\end{equation}
where $t$ is the tunneling amplitude.

We will solve this problem using the standard bosonization 
approach~\cite{bosonization}.  
The right and left moving chiral Fermi fields are bosonized 
according to the Mandelstam 
formulas
\begin{equation}
\psi_\pm^\dagger(x)\propto \frac{1}{\sqrt{2\pi}}
e^{\D\pm i \phi_\pm (x)}
\label{eq:mandelstam}
\end{equation}
where $\phi_\pm$ are chiral right and left moving bose fields respectively.
In the notation of ref.\ ~\cite{chamon}, the Lagrangians for the decoupled edges are
\begin{equation}
{\mathcal L}_\pm[\phi_\pm]=\frac{1}{4\pi} 
\partial_x \phi_\pm (\pm  \partial_t -v_0 \partial_x) \phi_\pm. 
\label{eq:Lpm}
\end{equation}
The normal-ordered
density operators are bosonized according to the rules
\begin{equation}
J_{\pm}=-{\D{\frac{1}{2\pi}}} \partial_x\phi_\pm.
\label{bosonization}
\end{equation}
In terms of the chiral boson fields $\phi_\pm$, the full 
(bosonized) Lagrangian density is 
\begin{eqnarray}
{\mathcal L}&=&\frac{1}{4\pi} 
\partial_x \phi_+ (  \partial_t -v_0 \partial_x) \phi_+ +
\frac{1}{4\pi} 
\partial_x \phi_- (- \partial_t -v_0 \partial_x) \phi_- 
\nonumber \\
&&-\frac{2g_c}{4\pi} 
\partial_x \phi_+ \partial_x \phi_- 
-\delta(x)\Gamma\D\cos\left(\phi_++\phi_-\right)
\nonumber \\
&&
\label{totH}
\end{eqnarray}
where $\Gamma$ measures the tunneling amplitude. As usual, this system is
diagonalized by the (Bogoliubov) transformation
\begin{eqnarray}
\phi_+&=&\frac{K+1}{2\sqrt{K}} \;
\varphi_++\frac{K-1}{2\sqrt{K}} \; \varphi_-
\nonumber \\
\phi_-&=&\frac{K-1}{2\sqrt{K}} \; \varphi_+
+\frac{K+1}{2\sqrt{K}} \; \varphi_-
\nonumber \\
&&
\label{bogoliubov}
\end{eqnarray}
and the choice of $K$ that diagonalizes the system is the effective Luttinger parameter.
Letting $v$ denote the renormalized velocity respectively, the
effective Luttinger parameter and the renormalized velocity are given
respectively by
\begin{equation}
K\equiv\sqrt{\D\frac{1-g_c}{1+g_c}}\qquad v\equiv
v_0\sqrt{1-g_c^2}
\label{K}
\end{equation}
With these definitions the Lagrangian density for the line junction with a point
contact at $x=0$ becomes
\begin{eqnarray}
{\mathcal L}&=&\frac{1}{4\pi} 
\partial_x \varphi_+ (  \partial_t -v \partial_x) \varphi_+ +
\frac{1}{4\pi} 
\partial_x \varphi_- (- \partial_t -v \partial_x) \varphi_- 
\nonumber \\
&& 
-\delta(x)\Gamma\D\cos\left[\smash{\sqrt{K}}(\varphi_++\varphi_-)\right]
\nonumber \\
&&
\label{Lfinal}
\end{eqnarray}
By comparison with ref.\ ~\cite{chamon} we see that the Luttinger parameter $K$ plays
the role of an effective inverse filling factor $\bar \nu=1/K$.
With the notation that we are using here $K$ plays the role of the constant 
$g$ defined in Ref.~\cite{kane}. 

\begin{figure}[!t]
\begin{center}
\leavevmode
%\vspace{.2cm}
\noindent
%\hspace{1.0 in}
\epsfxsize=.30\textwidth
\epsfbox{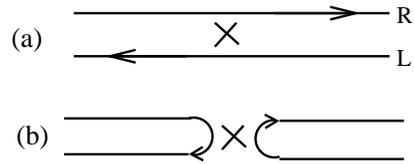}
%\vspace{.5cm}
\end{center}
\caption{Two phases for the system described by the Hamiltonian density
  Eq.~(\ref{Lfinal}) depending on the value of the Luttinger parameter
  $K$ defined by Eq.~(\ref{K}) at $T=0$. There is  a quantum phase
  transition at $K=1$ between two phases~\cite{kane}: (a) a perfectly conducting regime  
  for $K>1$, which corresponds to no tunneling in our problem, and (b) a perfectly 
  insulating
  regime for  $K<1$, which corresponds to the perfect
  tunneling in  our problem}
\label{fig:kane}
\end{figure}

Kane and Fisher studied transport properties of a one-dimensional
electron gas with a single impurity in Ref.~\cite{kane}
and predicted a change in the nature of the transport across the point contact 
(the impurity) at $T=0$ depending on the
value of Luttinger parameter $K$, finding perfect transmission for $K>1$
and  perfect insulating behavior for $K<1$ due to a complete backscattering
at the impurity, see Fig.~\ref{fig:kane}. 

%%%%%%%%%%%%%%%%%%%%%%%%%%%%%%%%%%%%%%%%%%%%%%%%%%%%%

%%%%%%%%%%%%%%%%%%%%%%%%%%%%%%%%%%%%%%%%%%%%%%%%%%%%%%%
As discussed in the caption of Fig.~\ref{fig:kane},  perfect conduction along the
wire in the Kane-Fisher problem ~\cite{kane,extra-kane} corresponds to conduction {\sl only}
along the barrier in our problem, and hence to the complete suppression of tunneling across 
junction in our case, and vice versa. 
However, we should keep in mind that the expression of
Eq.~(\ref{K}) can be the correct expression for the Luttinger
parameter only when the dimensionless coupling constant $g_c$ in Eq.~(\ref{HG}) 
is small. From our previous estimation of $g_c$ we found that 
to lowest order in $U/E_F$, $g_c$ is substantially large. Hence the effects of
irrelevant operators not included in the Hamiltonian $H$ cannot be 
ignored as they
will give rise to finite, and presumably not small, corrections to the functional 
dependence of the Luttinger
parameter $K$ on dimensionless Coulomb interaction $g_c$. Nevertheless, what matters 
is that even after all these corrections are accounted for there is an effective 
Luttinger parameter $K$, albeit with a complicated but analytic dependence on 
microscopic parameters. Thus 
we can still define an effective coupling constant $\tilde g_c$ through an identity 
of the form
$K\equiv \sqrt{\frac{1-{\tilde{g_c}}}{1+{\tilde{g_c}}}}$, where 
$\tilde g_c(\nu)=f(g_c(\nu))=g_c(\nu)+O(g_c^2)$. Therefore
all we can tell from Eq.~(\ref{K}) is that the Luttinger parameter
will become substantially smaller than $1$ due to Coulomb
interaction effects. However what matters here is that this condition is sufficient 
to bring the junction deep into
the back scattering phase at zero temperature where back
scattering is a strongly relevant perturbation. In this regime the perturbative
approach of Kane and Fisher~\cite{kane} is not enough to determine 
the transport properties at finite temperature. Fortunately,
this problem can be mapped to an exactly solvable boundary sine-Gordon (BSG)
problem which will enable us to go beyond the perturbative regime. We
analyze the problem from the perspective of BSG in the next section.

%%%%%%%%%%%%%%%%%%%%%%%%%%%%%%%%%%%%%%%%%%%%%%%%%%%%%%%%%%%%%%%%%%%%%%%%%%%%%%%%%%%%%
\section{Tunneling conductance near $\nu=1$}
\label{sec:BSG}
%%%%%%%%%%%%%%%%%%%%%%%%%%%%%%%%%%%%%%%%%%%%%%%%%%%%%%%%%%%%%%%%%%%%%%%%%%%%%%%%%%%%%
%%%%%%%%%%%%%%%%%%%%%%%%%%%%%%%%%%%%%%%%%%%%%%%%%%%%%%%%%%%%%%%%%%%%%%%%%%%%%%%%%%%%
\subsection{Mapping to the boundary sine-Gordon model}
%%%%%%%%%%%%%%%%%%%%%%%%%%%%%%%%%%%%%%%%%%%%%%%%%%%%%%%%%%%%%%%%%%%%%%%%%%%%%%%%%%%

In order to make contact with the results of Fendley, Ludwig and Saleur, we will now map the
effective Lagrangian of Eq.\ (\ref{Lfinal}) to the boundary sine-Gordon (BSG) theory. To that
effect we will perform a parity operation $x \to -x$ acting only 
on the left moving field $\varphi_-$ by which it now
becomes a right moving chiral boson, still denoted by $\varphi_-$. Let us define the even and
odd linear combinations of (right moving) chiral fields
\begin{eqnarray}
\varphi_e &=&\frac{1}{\sqrt{2}}(\varphi_++\varphi_-)
\nonumber \\
\varphi_o&=&\frac{1}{\sqrt{2}}(-\varphi_++\varphi_-)
\nonumber \\
&&
\label{even-odd}
\end{eqnarray}
in terms of which the Lagrangian takes the simpler, decoupled, form
\begin{eqnarray}
{\mathcal L}&=&\frac{1}{4\pi} 
\partial_x \varphi_e (  \partial_t -v \partial_x) \varphi_e +
\frac{1}{4\pi} 
\partial_x \varphi_o ( \partial_t -v \partial_x) \varphi_o 
\nonumber \\
&& 
-\delta(x)\Gamma\D\cos\left[\smash{\sqrt{2K}}\varphi_e)\right]
\nonumber \\
&&
\label{Ldecoupled}
\end{eqnarray}
In terms of the right moving chiral bosons $\varphi_e$ and $\varphi_o$, the edge currents
$J_\pm$ become
\begin{eqnarray}
J_+&=&+\frac{K}{2\pi \sqrt{2K}} \;
\partial_x \varphi_o -
\frac{1}{2\pi\sqrt{2K}} \; 
\partial_x \varphi_e  
\nonumber \\
J_-&=&+\frac{K}{2\pi \sqrt{2K}} \;
\partial_x \varphi_o +
\frac{1}{2\pi\sqrt{2K}}  \;
\partial_x \varphi_e  
\nonumber \\
&&
\label{Jdecoupled}
\end{eqnarray}
In the presence of the point contact, the current along the junction splits into a 
back-scattering or tunneling current and a forward-scattering or transmitted current.
The tunneling current $J_t=J_+-J_-$ is given by
\begin{equation}
J_t=-\frac{1}{2\pi} \sqrt{\frac{2}{K}} \;
\partial_x \varphi_e 
\label{tunnel-current}
\end{equation}
and it depends only on the chiral boson $\varphi_e$.

In order to map the problem to the boundary sine-Gordon theory we will use the
standard folding procedure~\cite{andrei,affleck,polchinski,fendley,saleur}. Let $x_0=vt$
denote a rescaled time coordinate and $x_1=x$.
We map each of left moving fields
defined on the whole line $\varphi_e$ and $\varphi_o$ to non-chiral fields
$\Phi_e$ and $\Phi_o$ defined on the half-line $x_1\geq 0$. These non-chiral fields 
 are decomposed into their right and left moving parts:
\begin{eqnarray}
\Phi_e(x_0,x_1)&=&\Phi_{e,-}(x_1+x_0)\!+\!\Phi_{e,+}(-x_1+x_0) \nonumber\\
\Phi_o(x_0,x_1)&=&\Phi_{o,-}(x_1+x_0)\!+\!\Phi_{o,+}(-x_1+x_0)\nonumber\\
&&
\label{Phi}
\end{eqnarray}
where the right moving parts of the $\Phi$ fields come from the $x_1>\!0$ parts of the 
$\varphi$
fields, and the left moving parts of the $\Phi$ fields come from the $x_1<\!0$ parts 
of the  $\varphi$
fields:
\begin{equation}
\begin{array}{lll}
\Phi_{e,+}(x)\!\equiv\!\varphi_e(x) &\quad\Phi_{o,+}(x)\!\equiv\;\;\varphi_o(x)
\quad\text{for}&\quad x_1>0\\
\Phi_{e,-}(x)\!\equiv\!\varphi_e(x)&\quad\Phi_{o,-}(x)\!\equiv\!-\varphi_o(x)
\quad\text{for}&\quad x_1<0
\end{array}
\end{equation}
with $\Phi_{e/o,+}\!=\!0$ for $x_1<\!0$ and $\Phi_{e/o,-}\!=\!0$ for
 $x_1>\!0$.
In terms of the $\Phi$ fields, the Lagrangian density on the whole line of
Eq.~(\ref{Ldecoupled}) is mapped onto a Lagrangian density on the half line $x_1 \geq 0$,
\begin{equation}
{\mathcal L}
=\frac{1}{8\pi} (\partial_\mu\Phi_{e})^2+\frac{1}{8\pi} (\partial_\mu\Phi_{o})^2
-\delta(x_1) \; \frac{\Gamma}{v} \cos\left(\sqrt{\frac{K}{2}}\Phi_e \right)
\label{PhiH}
\end{equation}
In Eq.~(\ref{PhiH}), the odd boson $\Phi_o$ remains free, simply
obeying 
%Dirichlet 
Neumann boundary conditions at the origin
$\Phi_o(x_1=0)=0$, and decouples. In contrast, the even field $\Phi_e$, 
which from now on will be denoted by
$\Phi$ for simplicity, has a non-trivial dynamics governed by the
Lagrangian density
\begin{equation}
{\mathcal L}=\frac{1}{8\pi}(\partial_\mu\Phi)^2
-\delta(x_1)\; \; \frac{\Gamma}{v} \cos(\!\D\sqrt{\frac{K}{2}}\Phi)
\label{bsG}
\end{equation}
defined for $x_1 \geq 0$.
The (even) field $\Phi$ and obeys Neumann boundary conditions at both $x_1\!=\!0$ and
$x_1\to\infty$.

 The action of Eq.~(\ref{bsG}) is known as the
boundary sine-Gordon model and is a well-studied integrable
quantum field theory~\cite{ghoshal}. It is a
theory of a free scalar field coupled to the vertex operator
${\mathcal O}\!=\!\exp(i\sqrt{K/2} \; \Phi(0,t))$ at the boundary. 
The main physical effect of the tunneling operator is to induce a 
flow of boundary conditions (BC) ~\cite{ludwig-affleck} at $x_1=\!0$: for
$\Gamma\!=\!0$ $\Phi$ obeys a Neumann BC's at $x_1=\!0$, whereas for
$\Gamma\to\infty$ $\Phi$ has a Dirichlet BC's at $x_1=\!0$. The ( boundary)
scaling dimension for the operator $\mathcal{O}$ at the weak
coupling fixed point $\Gamma\to 0$ is
$d_{\mathcal{O}}\!=2 \left(\sqrt{K/2}\right)^2=\!K$. Thus for
$K\!<\!1$, as in our case, 
the tunneling operator is relevant and the weak coupling fixed point is
unstable. Conversely, in this regime the strong coupling fixed point is stable.
On the other hand, for  $K\!>\!1$, ${\mathcal O}$ is irrelevant at the weak
coupling fixed point and the system is more appropriately
described by a dual picture as in the case discussed in Ref.~\cite{chamon}. 
This is the conventional situation in the fractional quantum Hall regime.
In our case, Coulomb interaction reduced the value of $K$ to be
smaller than $1$ leading to a situation similar to the one considered by 
Fendley and coworkers~\cite{fendley}, who investigated the problem of
inter-edge  
quasi-particle tunneling in a FQH state.

We note in passing that in general, as noted in ref. ~\cite{fendley}, $4k_F$ processes
should be fine tuned to zero if $1/9 < K < 1/4$ for the system to be integrable. 
(This is so because
only one relevant perturbation is allowed for integrability~\cite{ghoshal}. ) 
Fortunately in the case of interest here $4k_F$ processes are
forbidden in a chiral system with only one tunneling center. 
Hence the system we are interested in is automatically fine tuned and the
problem is integrable even for $K<1/4$.

The (massless) boundary sine-Gordon theory, regarded as the massless 
limit of the conventional bulk sine-Gordon theory, was shown to be integrable by 
Zamolodchikov and Ghoshal~\cite{ghoshal}, who also determined the spectrum of the 
BSG system by means of the Thermodynamic Bethe Anstaz (TBA) 
for an arbitrary value of the Luttinger parameter $K$. The
spectrum contains a kink and an anti-kink and $n-2$ breathers for
$n-1<1/ K \leq n$. The case $K=1/2$ is special in that  there is
no breather and the even boson theory can be represented in terms of a free fermions. In
this case, kinks and anti-kinks are just particle-hole transforms
of ordinary fermions. Although this problem is solvable for any
value of $K$, the TBA computation is much simpler for
$1/K=m$ , where $m$ is an integer (in this case the bulk scattering matrix
is completely diagonal.)  Since we are interested in the regime $K<1$, 
we will focus in what follows on the case $K=1/m$, with integer $m$. 

In the problem of transport through a point contact with  $1/K$ integer there is a 
dynamically generated
scale $T_K$ which uniquely determines the low-energy
physics~\cite{fendley,chamon,fradkin}.( In this problem $T_K$ plays a role  similar 
to the Kondo temperature in the conventional
Kondo Problem of a magnetic impurity in a metallic host.) The scale $T_K$ is a function
 of the point-contact interaction strength
$\Gamma$ and of the ultra-violet cutoff scale $\Lambda$. $T_K$ is an energy scale 
separating the low energy, 
long distance regime (IR regime), and the high energy short distance regime (UV regime);
$T_K$ can also be viewed as the temperature at which 
the weak coupling expansion breaks down. One of the fundamental
properties of quantum impurity problems like point contact tunneling or the Kondo
model is that observables, such as  the differential conductance in the point contact 
problem or the magnetic susceptibility 
in the Kondo problem, are
described in the scaling regime by universal scaling functions of the temperature $T$, 
the bias voltage $V$ ($H/T$ for the Kondo
model) the coupling constant $\Gamma$ and the (ultraviolet) cutoff $\Lambda$, 
of the form
\begin{equation}
G(\Lambda,V,T,\Gamma)\stackrel{\longrightarrow}
{\scriptscriptstyle{T,V\ll\Lambda}}G(T/T_K,V/T)
\end{equation}
where the dependence of conductance upon cutoff and interaction strength is
hidden in the definition of $T_K$~\cite{saleur,andrei}. 
Fendley and coworkers~\cite{fendley} find a dependence of  $T_K$ on
$\Gamma$ of the form
\begin{equation}
T_K\!=\!C\Gamma^{\frac{1}{1-K}}, \label{TK}
\end{equation}
where $C$ is a non-universal constant. 

The rest of this section will be devoted to an analysis of the implications of the 
known results for the BSG model 
to the tunneling contact problem that we are interested in, and to its implications for the 
experiment of Kang and coworkers~\cite{kang}.
It will be shown that both Coulomb interaction and finite temperature
play important role in the behavior of the zero-bias conductance peak near $\nu\sim1$. 

%%%%%%%%%%%%%%%%%%%%%%%%%%%%%%%%%%%%%%%%%%%%%%%%%%%%%%%%%%%%%%%%%%%%%%%%%%%%%%%%%%%
\subsection{Comparison with the experiment}
%%%%%%%%%%%%%%%%%%%%%%%%%%%%%%%%%%%%%%%%%%%%%%%%%%%%%%%%%%%%%%%%%%%%%%%%%%%%%%%%%%%

We have shown above that the problem of the point contact in two laterally coupled FQH liquids
maps onto the boundary sine-Gordon theory. In particular we showed that the effective Luttinger
parameter $K$ plays the role of an effective inverse filling factor. In this picture the point
contact maps onto the problem of tunneling of electrons between two edges with filling factor
$\bar \nu=1/K>1$. Fendley, Ludwig and Saleur (FLS)~\cite{fendley} solved a very similar problem 
but 
in the regime $\bar \nu<1$. FLS also found that, at $T=0$ and voltage
$V$, the tunneling current $I$ obeys the exact 
remarkable duality
\begin{equation}
I(T_K,V,\bar \nu)=\frac{e^2}{h} \bar \nu V-\bar \nu^2 I(T_K,V,{\bar \nu}^{-1})
\label{duality}
\end{equation}
Using this result we find that the differential tunneling
conductance at zero temperature for our problem is given by~\cite{fendley,weiss}
\begin{equation}
G_t\!=\!K\frac{e^2}{h}\!\times\!
 \left\{\!\!\begin{array}{ll}
  \!1\!-\!\sum_{n=1}^\infty c_n(K^{-1})
  \left(\!\frac{eV}{T_K}\!\right)^{2n(K^{-1}\!-\!1)}\!&\!\frac{eV}{T_K}<e^\delta\\
 \!\sum_{n=1}^\infty c_n(K)\left(\frac{eV}{T_K}\right)^{2n(K\!-\!1)}
 &\frac{eV}{T_K}>e^\delta
 \end{array}\right.\label{zeroT}
\end{equation}
where the coefficients $c_n$ are defined as
\begin{equation}
c_n(K)\!=\!(-1)^{n\!+\!1}
\frac{\Gamma(nK\!+\!1)}{\Gamma(n\!+\!1)}\frac{\Gamma(1/2)}{\Gamma(n(K\!-\!1)\!+\!1/2)}
\end{equation}
where  $\Gamma(z)$ is the gamma function. 
( Here $\delta\!=\![K\ln K\!+\!(1\!-\!K)\ln(1\!-\!K)]/[2(1\!-\!K)]$ is a parameter that 
determines the radii of convergence of these series.)

\begin{figure}[!t]
\begin{center}
\leavevmode
%\vspace{.2cm}
\noindent
%\hspace{1.0 in}
\epsfxsize=.40\textwidth
\epsfbox{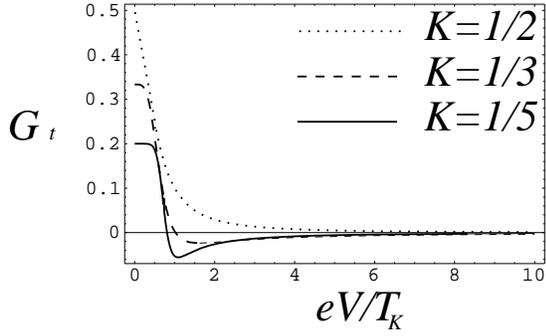}
%\vspace{.5cm}
\end{center}
\caption{The differential tunneling conductance at zero temperature v.\@s.\@ $eV/T_K$.
Each of dotted, dashed and solid line represents
    $K\!=\!1/2$,$K\!=\!1/3$ and $K\!=\!1/5$ respectively.
All three curves share the common feature of rapid increase in the 
$G_t$ as the voltage is
lowered past $T_K$ followed by the saturation of $G_t$ 
to the value determined by
Luttinger parameter $Ke^2/h$ at $V/T_K=0$.}

\label{fig:zeroT}
\end{figure}

In Fig.~\ref{fig:zeroT} we plot 
$G_t$ at zero temperature for different values of $K$ as
functions of $eV/T_K$ (in units of $e^2/h$).
%%%%%%%%%%%%%%%%%%%%%%%%%%%%%%%%%%%%%%%%%%%%%%%%%%%%%%%%%
%%%%%%%%%%%%%%%%%%%%%%%%%%%%%%%%%%%%%%%%%%%%%%%%%%%%%%%%%%
We can see from the plot that the
differential conductance increases rapidly as the voltage is
lowered below $T_K$ and that it saturates rather rapidly to a value determined by
Luttinger parameter $Ke^2/h$ at $V/T_K=0$ at zero temperature.
Recall that Eq.~(\ref{zeroT}) is valid only for $K<1$ where the
vertex operator is relevant~\cite{zwerger}. 
Thus, Eq.~(\ref{K}) implies that there exists a critical filling factor 
$\nu_c(g_c)$ for which $K=1$. For
$\nu>\nu_c$ we have $K<1$ since for repulsive Coulomb interactions $g_c>0$ and $\nu_c<1$. 
Therefore when the filling factor is increased past $\nu_c$, the vertex
operator becomes relevant and the tunneling amplitude $\Gamma$ flows to infinity (which
makes $T_K$ grow to infinity as well), leading to a finite
conductance at all bias voltages at zero temperature.
 
For values of the Luttinger parameter $K<1/2$ the tunneling conductance $G_t$, shown in 
Fig.~\ref{fig:zeroT} for $K\!=\!1/3$ and $K\!=\!1/5$, becomes
negative for sufficiently large values of $eV/T_K$. 
To understand this interesting feature we recall that the 
expression of tunneling current for $eV/T_K>e^\delta$ can be obtained from 
the second line of Eq.~(\ref{zeroT}) in the form
\begin{equation}
I_t(V)=\frac{e^2V}{h}K\sum_{n=1}^{\infty}a_n(K)\left(\frac{eV}{T_K}\right)^{2n(K-1)}
\label{IzeroT}
\end{equation}
with the coefficients $a_n$ given by
\begin{equation}
a_n(K)=\frac{1}{1/2+n(K-1)}c_n(K).
\end{equation}
Since the tunneling coupling should make the tunneling current increase, one
expects $a_1>0$ which implies $c_1(K)<0$
for $K<1/2$ from the above relation
between $a_n$ and $c_n$. 
This negative value of $c_1$ for $K<1/2$ causes the conductance to become negative at large
voltages,
and produces a dip in the conductance curve for $K=1/3$ and $K=1/5$ in
Fig.~\ref{fig:zeroT}.
This phenomena has same origin as the conductance along the quantum
wire becoming larger than $Ke^2/h$ in Ref.~\cite{fendley} and
Koutouza, Siano and Saleur reported similar phenomena in their work
where they considered charging effect on tunneling between quantum wires~\cite{koutouza}
However, the
negative conductance is expected only for practically infinite driving
voltage at zero temperature 
since $T_K$ is infinitely large at strong coupling fixed point and
numerical calculation of TBA shows that this effect disappears for
small $V/T$.~\cite{fendley}
    
Now let us turn to the finite temperature case.  In contrast to the
zero temperature behavior of indefinite running, 
the effective tunneling coupling $\Gamma$ stops 
running at a certain value  $\Gamma^*(T)$ determined by the
temperature at finite temperature. As in all quantum phase 
transitions~\cite{subir}, this effect in turn leads to the appearance of a finite 
temperature-dependent crossover scale $T_K^*(\Gamma^*(T),K)$. 
Furthermore, now both temperature and external
voltage act as natural crossover energy scales. From the point of view of our
scenario, finite temperature
 plays important role in understanding 
 the peculiar features of
the experiment, which can be summarized as follows:
\begin{enumerate}
\item Existence of region in filling factor with ZBC peak.
\item Substantially low height of conductance peak as comparison to typical
  Hall conductance $\nu e^2/h$.
\item Appreciably large width of the peak region beginning at $\nu^*\sim 1$.
\item Disappearance of the ZBC peak as $\nu$ is increased beyond $\nu\sim1$.
\item Reappearance of the ZBC peak in a region near and above  $\nu\sim2$.
\end{enumerate}
It turns out that, except for the reappearance of the zero-bias peak near $\nu=2$, 
most of these effects can be 
understood within the point contact scenario that we advocate here provided thermal 
crossover effects are taken into account.
The reappearance of the peak near
$\nu\sim2$  will be discussed in the next section. The rest of this
section will be devoted to our understanding on the first four aspects.

A central feature of this problem is the powerful fact that the differential tunneling conductance
$G_t$ is a universal scaling function of two dimensionless ratios, $T/T_K$ and $V/T$. First of all,
the system  behaves qualitatively as if it were at $T\!=\!0$ so long
as the temperature is the smallest among three energy scales i.\ e.\ , $T\ll T_K,V$. In this
regime the system flows to the stable fixed point at $\Gamma \to
\infty$ where
the tunneling current is large and the conductance saturates to its largest value 
$K\frac{e^2}{h}$ at ZBC. However, since the crossover scale $T_K$ is a (weak) function of
$\nu$, there exists a filling factor $\nu^*$ for which $T_K(\nu^*)=T_K^* \sim T$. For $T>T_K^*$
the system will flow toward the decoupled unstable fixed point at $\Gamma=0$. Hence,
in contrast with the case $T=0$ we expect only a crossover, instead of a phase transition. 
In particular this also means that, at low but fixed temperature $T$, we should
see an appreciable {\sl increase} in $G_t$ when $V$ becomes smaller
than $T_K^*$, since $T_K^*$ will be finite at non-zero temperature. 
However, as $V$ becomes comparable to $T$ the system will
begin to be driven by thermal fluctuations, and the coupling $\Gamma$ would no 
longer increase further as the voltage is lowered, thus leading to a saturation of the tunneling
conductance at a value determined by temperature. Therefore, even though the
ZBC peak should be observable due to an increase in $G_t$ as the voltage is
lowered past $T_K^*$, the height of the peak (essentially determined by the temperature) would
be much lower than the zero temperature saturation value $Ke^2/h$.
Conversely, if the temperature is higher than $T_K$, thermal fluctuations 
dominate for all values of
$V$ and no ZBC peak should be observed. 

On the other hand, in the regime where the filling factor is such that $K<1$, the dependence
of $T_K$ on the tunneling amplitude $\Gamma$ is such (see Eq.~\ref{TK}) that
as  the filling factor 
$\nu$ increases, the exponent in the dependence of $T_K$ upon
$\Gamma$ decreases. Hence, as $\nu$ is increased well past a value $\nu\sim1$, 
the crossover scale
$T_K$ decreases, and at some point it becomes lower than the temperature. In this regime the
junction is effectively in the high temperature regime and the ZBC peak 
is absent. Thus, in the point contact scenario, the gradual but rapid disappearance 
of the ZBC peak is a manifestation of this crossover.

This discussion can be made more explicit by looking at the behavior of the
$\beta$-function defined as
\begin{equation}
\beta(\Gamma,V,T)\!\equiv\! -\frac{\partial\Gamma}{\partial\ln V},
\label{def_beta}
\end{equation}
This renormalization group function measures the change of the effective coupling constant 
$\Gamma$ at temperature $T$ as the external
voltage $V$ is varied.
The statement that the conductance is a scaling function of the ratios $T/T_K$ and $V/T$,
is equivalent to say that one can define  a set of systems which 
have the same conductance 
as the external voltage is varied. This set of equivalent systems amounts to a renormalization
group flow defined by the Callan-Symanzik equation
\begin{equation}
\frac{dG_t}{d\ln V}(T/T_K,V/T)\!=\!\frac{\partial G_t}{\partial\ln
  V}\!+\!\frac{\partial\Gamma}{\partial\ln V}\frac{\partial G_t}{\partial\Gamma}=0,
\label{callan}
\end{equation}
where the second term on the right hand side of the first equality comes from
the fact that $T_K$ has intrinsic dependence upon the coupling constant
$\Gamma$. Note that in Eq.~(\ref{callan}) 
 we chose to vary the energy scale $V$ instead of the cut off scale,
which as usual is hidden in the definition of $T_K$. This equation can be used
to calculate the $\beta$-function defined in Eq.~(\ref{def_beta}):
\begin{equation}
\beta(\Gamma,V,T)\!=\!\frac{\frac{\partial G_t}{\partial\ln V}}{\frac{\partial
    G_t}{\partial\Gamma}}\!=\!\frac{V\frac{\partial G_t}{\partial
    V}}{\frac{1}{1\!-\!K}\frac{T_K}{\Gamma}\frac{\partial G_t}{\partial T_K}},
\label{beta}
\end{equation}  
where we used the relation between $T_K$ and $\Gamma$ Eq.~(\ref{TK}) for
the second equality.

At zero temperature, using Eq.\ ~(\ref{zeroT}) it is easy to see that 
 $V\partial G_t/\partial V=
-T_K\partial G_t/\partial T_K$, and we obtain the expected result~\cite{zwerger}
\begin{equation}
\beta(\Gamma,V,T\!=\!0)\!=\!-(1\!-\!K)\Gamma,
\label{T_zero_limit_beta}
\end{equation}

In order to analyze the $\beta$-function at finite temperature, we now 
turn to $K\!=\!1/2$ case in which exact $G_t$ is known in closed 
form even at 
finite temperature, by refermionizing the even boson theory to
non-interacting spinless free fermion. In this special case, not
only the conductance but all $n-$point correlation functions are
exactly solvable~\cite{kane,fendley,chamon_freed} and the integral
in the Eq.~(5.2) of Ref.~\cite{fendley} can be reexpressed in
terms of the digamma function $ \psi(x)\!=\!\Gamma'(x)/\Gamma(x)$
leading to the expression for the conductance
\begin{equation}
G_t(T,V,K\!=\!1/2)=\frac{1}{2}\frac{e^2}{h}\frac{T_K}{\pi
T}Re\psi'\left[\frac{1}{2}\!+\!\frac{T_K}{\pi T}\!+\!\frac{ieV}{2\pi
T}\right].
\label{eq:Gexact}
\end{equation} 
%%%%%%%%%%%%%%%%%%%%%%%%%%%%%%%%%%%%%%%%%%%%%%%%%%%%%%%%%%%%%%%%%%
\begin{figure}[h!]
\begin{center}

\epsfig{file=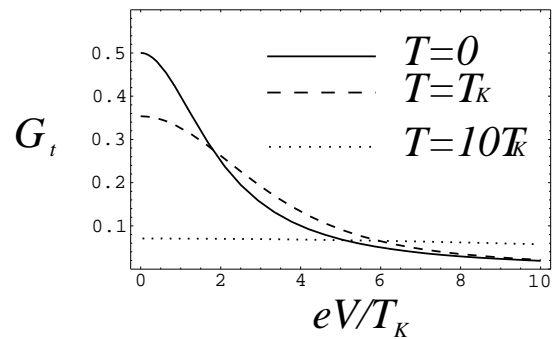,width=.4\textwidth}\par

\par
%\end{center}
 \caption{ The exact differential tunneling conductance given by
    Eq.~(\ref{eq:Gexact}) is plotted  
    as a function of $eV/T_K$ for different values of $T/T_K$ for $K\!=\!1/2$. 
    Observe the lowering and broadening 
    of the peak as the temperature is increased.}
\vspace{.5cm}
%\psfrag{x}{$T/T_K$}\psfrag{y}{$G_t(V=0)$}

%(b)\par
%\vspace{.5cm}
\end{center}
\label{fig:exact3D-all}
\end{figure}
%%%%%%%%%%%%%%%%%%%%%%%%%%%%%%%%%%%%%%%%%%%%%%%%%%%%%%%%%%
\begin{figure}[h!]
\begin{center}
\epsfig{file=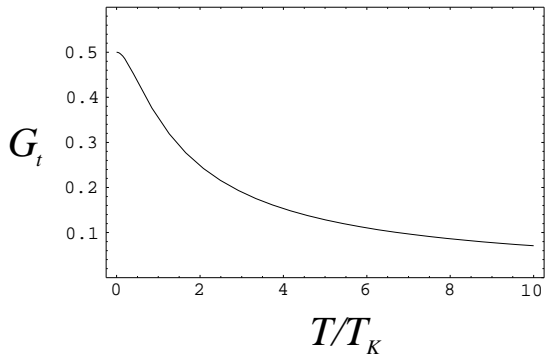,width=.4\textwidth}\par
 \caption{ The zero bias conductance peak
    height, $G_t(T,V=0,K\!=\!1/2)$ is plotted as a function of temperature.}
    \end{center}
 \label{fig:exact3D}
 \end{figure}
%%%%%%%%%%%%%%%%%%%%%%%%%%%%%%%%%%%%%%%%%%%%%%%%%%%%%%%%%%%%%%%%%%%%%

The plot of $G_t(T,V,K\!\!=\!\!1/2)$ as a function of $eV/T_K$ for
several values of $T/T_K$ 
in Fig.~\ref{fig:exact3D-all} shows the broadening of the peak as the
temperature is increased. 
The reduction and eventual disappearance of peak height at high
temperatures is quite obvious in the plot of ZBC peak
height as a function of $T/T_K$
in Fig.~\ref{fig:exact3D}. 
One can also understand the gradual disappearance of
ZBC peak as $\nu$ is further increased as following. From
Eqs.~(\ref{TK}) and (\ref{K}) , we see that $T_K$ decreases as
$\nu$ increases for $K\!<\!1$. Therefore as $\nu$ becomes larger at
given $T$, $G_t$ will be determined by lower $T_K$ leading to
smaller ZBC peak which would eventually disappear.

%%%%%%%%%%%%%%%%%%%%%%%%%%%%%%%%%%%%%%%%%%%%%%%%%%%%%%%%%%%%%%%%%%%%%%%%%
\begin{figure}[t!]
\begin{center}
\leavevmode
%\vspace{.2cm}
\noindent
%\hspace{1.0 in}
\epsfxsize=.40\textwidth
\epsfbox{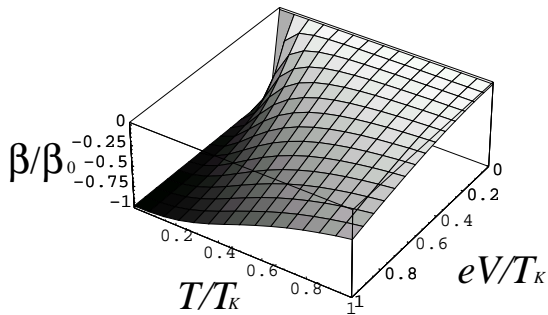}
\end{center}
\caption{The exact $\beta$-function in units of
$\beta_0\!\equiv\!\beta(V,T\!=\!0)\!=\!\Gamma/2$ is shown as a function of
$eV/T_K$ and $T/T_K$. There is a
cross-over between $T\!\to\!0, V\!\neq\!0$ limit, where
$\beta\!\to\!-1/2\Gamma$ and $V\!\to\!0, T\!\neq\!0$ limit, where
$\beta\!\to\!0$ near $T\!\sim\! V$: as the temperature is increased past
$V$, the $\beta$-function approaches zero where the coupling stops to
run. This cross-over explains the low height of peak, which eventually 
disappears as $\nu$ is increased well beyond 1 leading to a smaller $T_K$.}\label{fig:beta}
\end{figure}
%%%%%%%%%%%%%%%%%%%%%%%%%%%%%%%%%%%%%%%%%%%%%%%%%%%%%%%%%%%%%%%%%%%%%%%%%

The role of the temperature on the peak height can also be seen 
by looking at the asymptotic behavior of $G_t(T,V\!=\!0,K\!=\!1/2)$ in the 
limit of $T\!\to\!0$.
At $V\!=\!0$, $G_t\!=\!\frac{1}{2}\frac{e^2}{h}\frac{T_K}{\pi
  T}\psi'(\frac{1}{2}\!+\!\frac{T_K}{\pi T})$ from Eq.~(\ref{eq:Gexact}).
In the limit
$T\!\to\!0$, we can use the asymptotic expansion of the digamma function
\begin{equation}
\psi(x)\!\sim\!\ln x\!-\!\frac{1}{2x}+ \ldots \; , \quad{\mathrm {for} }\;\; |x|\gg1
\label{asymp_digamma}
\end{equation}
to infer the asymptotic behavior of peak height
in the low temperature limit as
\begin{equation}
G_t\sim\frac{1}{2}\frac{e^2}{h}\left[1\!-\!\frac{1}{4}\left(\frac{\pi T}{T_K}\right)^2-
\cdots\right],
\end{equation}
where the decrease of peak height at finite temperature is evident. 

Although it is possible to calculate the differential conductance at zero bias 
for more general values of the Luttinger parameter $K$, it involves solving a set of complex
coupled integral equations. This has been done numerically for the related problem 
of tunneling into a
Luttinger liquid in the work of  Koutouza, Siano and Saleur~\cite{koutouza} 
who find that the results vary quite smoothly as $K$ changes below $1/2$. (The main differences
arise due to an analog of the ``resonance" found earlier by Fendley, Ludwig and 
Saleur~\cite{fendley}. This resonance is responsible for the negative differential conductance at
large voltages and at $T=0$.)) Thus, at least at a
qualitative level, it seems that the behavior for $K$ below $1/2$ can be described by a curve like
that of Eq.\ \ref{eq:Gexact}, for some crossover scale $T_K$, but with $K$ replacing the overall 
factor of $1/2$. Preliminary results indicate that this is also a quantitaively accurate
description of the data~\cite{kang-priv}. 

With the full expression for the conductance Eq.~(\ref{eq:Gexact}),
we can calculate the beta function Eq.~(\ref{beta}) to obtain
\begin{equation}
\beta(\Gamma,V,T)\!=-\frac{1}{2}\Gamma\frac{eV}{2\pi
    T}\frac{{\mathrm 
    Im}\psi^{(2)}(z)}{{\mathrm
    Re}\psi^{(1)}(z)\!+\!\frac{T_K}{\pi T}{\mathrm Re}\psi^{(2)}(z)},
\label{exact-beta}
\end{equation}
where $z=1/2\!\!+\!\!T_K/\pi T\!\!+\!\!ieV/2\pi T$ and $\psi^{(n)}(z)$ 
stands for $n$'th derivative of the digamma function.
This result is shown in Fig.~\ref{fig:beta} in the form of 
the plot of $\beta(V,T)/|\beta(V,T\!=0\!)|$ as a function
of $T/T_K$ and $eV/T_K$. 
~From the above expression, we can immediately read off that 
\begin{equation}
\lim_{V\!\to\! 0}\beta(\Gamma,V,T\!\neq\!0)=0,
\label{V_zero_limit_beta}
\end{equation}
which means that the coupling stops running at $V\!=\!0$ at finite 
temperature. 
Comparing Eq.~(\ref{V_zero_limit_beta}) to Eq.~(\ref{T_zero_limit_beta}) which
gives $\beta(\Gamma,V,T\!=\!0)\!=\!-\Gamma/2$ for the case of 
consideration $K\!=\!1/2$, we can see that the limits $T\!\to\!0$ and 
$V\!\to\!0$ do not commute. Hence, we conclude that there is a singularity at $T\!=\!V\!=\!0$, 
simply illustrating the fact that the coupling runs indefinitely 
only at zero temperature due to the underlying
quantum phase transition at $K\!=\!1$. 
This implies that all 
we should be able to see at any finite temperature is be a 
crossover from $T\!>\!T_K$ to $T\!<\!T_K$ near
$K\!\sim\! K^*\!<\!1$ at which the tunneling
increase rapidly as $K$ becomes smaller than $K^*$, giving rise to
a pronounced ZBC peak as $\nu$ is increased pass
$\nu^*\!>\!\nu_c$. This explains why the experiment sees a rapid
increase of the ZBC peak when $\nu$ is increased past
$\nu^*\!\gtrsim\!1$, even though we expect $\nu_c\!<\!1$ due to the effects of the Coulomb
interaction. 
Furthermore, since the behavior of $\beta$-function is quite 
different for $V\!=\!0$ and $T\!=\!0$, we expect a crossover near 
$V\!\sim\! T$, which was discussed earlier in relation to 
the existence of peak region with finite width and 
the low height of the peak in the region. These crossover effects, and the behavior of the
beta-function, are shown in Fig.~\ref{fig:beta}.
This result illustrates our general statement that  
it is the competition 
between the temperature and the bias voltage what
enables us to observe the conductance peak,
and the height of the peak can  be much
lower than the saturation value $Ke^2/h$ since the observable height will be
limited by the temperature. This result also
supports our argument that competition between 
$T$ and $T_K$ eventually leads to the disappearance
of peak as the filling factor is raised further past a value $\nu\!\sim\!1$.

%In Fig.\ \ref{fig:fit} we present a fit of data from W.\ Kang and coworkers~\cite{kang-new} on the
%temperature dependence of the zero bias conductance peak in the temperature range from 
%$T=300 \; mK$ 
%to $ 10\; K$. The data is consistent with an effective Luttinger parameter $K \sim 1/6$. Instead
%of using the full Bethe-Ansatz result for $K=1/6$ we have chosen to fit the data by means of 
%the (simpler) function $G(T,V=0)$ of Eq.\ (\ref{eq:Gexact}), which is exact for $K=1/2$, but 
%changing the overall scale from $1/2$ to $K\sim 1/6$. While this procedure is not rigurous, it
%should be accurate at low temperatures. 
%The result of the fit, shown in Fig.\
%\ref{fig:fit}, lends strong support to our interpretation of this experiment~\cite{comment}.

In this section, we gave a detailed analysis of the experimental predictions of
our model, which was developed in the previous section.
After mapping the problem to BSG model, by
borrowing known exact results of BSG problem and 
calculating the relevant $\beta$-function,
we suggested consistent explanations to so far not understood
peculiar features of the experiment by Kang and coworkers.
In our picture, finite temperature effects are responsible for 
the observability of a ZBC peak with an unexpectedly low value of its height, as well as to 
the finite width in
filling factors where the peak is observed. 
%%%%%%%%%%%%%%%%%%%%%%%%%%%%%%%%%%%%%%%%%%%%%%%%%%%%%%%%%%%%%%%%%%%%%%%%%
%\begin{figure}[h!]
%\begin{center}
%\epsfig{file=fitdata2.eps, width=.4\textwidth}
%\end{center}
%\caption{Comparison between experimental and theoretical ZBC peak
%height at  finite temperature. The theoretical fit to the data was done
%using Eq.~(\ref{eq:Gexact}) at $V=0$ with the overall factor $K=1/2$ replaced by
%$K=0.1685 \sim 1/6$ and $T_K=4.5634\; K$. (See discussion in the text and in
%footnote~\cite{comment}.) (The unpublished
%experimental data is courtesy of W.\ Kang.)}
%\label{fig:fit}
%\end{figure}
%%%%%%%%%%%%%%%%%%%%%%%%%%%%%%%%%%%%%%%%%%%%%%%%%%%%%%%%%%%%%%%%%%%%%%%%%

Our picture is a natural consistent scenario for the
appearance of the ZBC peak in the wide range of 
filling factor near $\nu\sim1$. In the next section we will show that the 
reappearance of the ZBC peak in the range of filling factor near
$\nu\sim2$ in principle can also be understood by following
closely the approach of this section for the $\nu\sim1$ case but now considering 
the possibility of a partially  spin-polarized state near $\nu\sim2$.

%%%%%%%%%%%%%%%%%%%%%%%%%%%%%%%%%%%%%%%%%%%%%%%%%%%%%%%%%%%%%%%%%%%%%%%%%%%%%%%%%%%%%%%
\section{Tunneling conductance near and above $\nu=2$}
\label{sec:spinful} 
%%%%%%%%%%%%%%%%%%%%%%%%%%%%%%%%%%%%%%%%%%%%%%%%%%%%%%%%%%%%%%%%%%%%%%%%%%%%%%%%%%%%%%%

To understand the peak in the  $\nu\sim2$ region, we first note that this peak region begins
abruptly near $\nu\sim2$ in apparently similar manner as it
does near $\nu\sim1$~\cite{kang}. In the bulk system, as the filling factor
becomes comparable to $\nu \sim 2$, the electron spin begins 
to matter, as the spin reversed states begin to get progressively occupied.
Thus, even if the 2DEG is fully polarized  for $\nu \sim 1$, spin plays a
crucial role for $\nu \sim 2$. In addition, for samples with high nominal
electronic density, spin fluctuations are known to become important and in some
cases so much that the ground state may even be a spin singlet. However, this situation 
requires samples with fairly high densities, which is presumably not the case
in the experiment of Kang and coworkers. Hence, a natural extension of the
picture that we advocated for in the previous section, as it stands applicable
only for fully polarized 2DEGs, simply requires to take into account 
the changes in the physics brought about by the
electron spin, and in particular of the role played by both
Zeeman and exchange interactions. This extension should be applicable to both 
spin singlet and non-singlet cases.
However, once the spin degree of freedom is included, there is a richer class of possible
behaviors, for there are now three possible types of tunneling corresponding 
to tunneling of charge and/or spin degree of freedom. In what follows
we will be interested mostly in the  
regime in which the spin polarization is not large. Hence, we will assume a 
reference state in which the up and down spin branches have the same filling factor
$\nu_\uparrow=\nu_\downarrow$, and investigate the effects of the 
Zeeman term which will tend to polarize the state. We will focus on states with total filling
factor $\nu\geq 1$. For these states the outermost edge is a $\nu=1$ edge (per spin component).
The effects of the magnetic field thus enters in the choice of the
range of filling factor $\nu \geq 2$, in the presence of spin exchange interactions, and in
the effects of the Zeeman  term as well as other possible $SU(2)$ symmetry breaking terms
on the edge states. We will consider two different physical
situations: 1) when the $SU(2)$ symmetry of spin is broken either by a (large) Zeeman term, 
in which case
the ground state may be polarized (although not necessarily fully polarized), or by magnetic
anisotropy terms (expected to be very small in these systems), and 2) when the Zeeman
term is small enough that the ground state is a singlet at $\nu=2$. There are a number of other 
interesting cases, such as the singlet and partially polarized states at $\nu<2$ which will not be
discussed here. These states have interesting tunneling properties~\cite{ana2} but do not exhibit
the ZBC peak in the tunneling conductance that we are discussing here. 

The Hamiltonian density that was studied in the previous section
can be easily modified to account for the spin degree of freedom and its interactions.
Thus, we write the Hamiltonian
density  in terms of the spin dependent chiral electron densities 
$J_{\pm,\alpha}\equiv \psi^\dagger_{\pm,\alpha} \psi_{\pm,\alpha}$, with spin projection 
$\alpha=\uparrow , \downarrow$. 
Furthermore spin-spin exchange interaction and the Zeeman term should now be included in the 
Hamiltonian density.
Let us define charge densities operators of chiral modes as $J_{\pm}^c\equiv
J_{\pm, \uparrow}+J_{\pm, \downarrow}$, and the three-component spin densities
operator  $J_\pm^{a}\equiv \frac{1}{2}
\psi_{\pm, \alpha}^\dagger\sigma^a_{\alpha\beta}\psi_{\pm, \beta}$, where
$\sigma^a$ are Pauli matrices, with $a=x,y,z$. The Hamiltonian density for 
the system of  two coupled edges can be written as a sum of charge and spin Hamiltonians, 
\begin{equation}
\mathcal H_G=\mathcal H_c+\mathcal H_s
\end{equation}
The charge Hamiltonian is given by
\begin{equation}
\mathcal H_c=\frac{\pi v_c}{2} \left(J_+^c J_+^c+J_-^c J_-^c + 2g_c J_+^c J_-^c\right)
\label{eq:Hc}
\end{equation} 
where both the bare edge velocity and 
the effects of the intra-edge interactions are absorbed in the
effective charge velocity $v_c$. We will write the spin part of the
Hamiltonian as a sum 
of two terms
\begin{equation}
\mathcal H_s=\mathcal H_{\rm symm}+\mathcal H_{\rm pert}
\label{eq:Hs}
\end{equation}
where $SU(2)$ invariant part has the form~\cite{spin}
\begin{equation}
\mathcal H_{\rm symm}=\frac{2\pi}{3} v_s \left(\vec J_+ \cdot \vec J_++\vec J_- \cdot \vec J_- +6g_s
\vec J_+ \cdot \vec J_-\right)
\label{eq:Hs-symm}
\end{equation}
Here $v_s$ includes the effects of intra-edge spin interactions and $g_s$ is the inter-edge
strength of the exchange interaction. 

We have used a simple and rather crude model to estimate the inter-edge exchange coupling constant. We
modelled the barrier with a potential $V(x)$ of height $V_0$ and width $2a$. We find that, as expected, 
due to the antisymmetry of the wave function the 
dimensionless coupling constant $g_s$ has ferromagnetic sign, and that its magnitude has a rapid
dependence of $k_F\ell$ where $k_F$ is the Fermi wave vector of the edge states. For 
a barrier of width $88 \AA$ and height $220 \; meV$, and for a model in which correleations enter only
in the antisymmetry of the wave function, we estimate that reasonable values of the 
dimensionless inter-edge exchange coupling constant are quite small, typically in the range 
$|g_s| \sim 10^{-3}$ to $10^{-4}$. While it is quite possible that we are underestimating the magnitude
of $g_s$ it seems unlikely that a realistic value can be larger by more that an order of magnitude.
In addition, we show below that for the ferromagnetic sign, inter-edge exchange interactions are
(marginally) irrelevant. Hence it is reasonable to set $g_s$ to zero if $g_s<0$, since the expected 
(logarithmic) corrections to scaling will be exceedingly small. 

The Hamiltonian for the symmetry breaking
perturbations, \textit{i.\ e.\/} a Zeeman term and an
anisotropy term, is
\begin{equation}
\mathcal H_{\rm pert}=-\mu_B g B \left(J_+^z+J_-^z\right)+4\pi v_s g_s \lambda J_+^z J_-^z
\label{eq:Hpert-spin}
\end{equation}
where $\mu_B$ is the Bohr magneton,  $g$ is the gyromagnetic ratio, and $\lambda$ measures the
strength of the magnetic anisotropy (which is  quiet likely to be very small in the samples of 
Kang and coworkers). For $\lambda >0$ the anisotropy is Ising like and for $\lambda <0$ it is $XY$
like. For notational convenience, we define $XY$ component exchange
coupling $g_\perp$ and Ising exchange coupling $g_\parallel$: 
\begin{equation}
g_\perp\equiv g_s, \qquad g_\parallel\equiv(1+\lambda) g_s. 
\label{eq:g-perp-g-par}
\end{equation}
Experimentally it is known that in $GaAs$ the gyromagnetic factor is anisotropic and that this
anisotropy is quite large for the geometry of the experiment of Kang and coworkers~\cite{awschalom}.
The magnitude (and sign) of the magnetic anisotropy (anisotropy in the exchange interaction) are
apparently not known. As we will see below magnetic anisotropy can potentially lead to 
 interesting effects such as a possible spin gap state. However given the smallness of our estimate of
 the exchange interaction we should expect the spin gap to be small as well. Nevertheless, in spite of
 the possible  small value of this gap, we will
 discuss the interesting physics of this state.

We will treat the spin-$1/2$ case using Abelian bosonization, much in the way we did the
spin-polarized case in the previous section. However, we will pay special attention to the role of
the $SU(2)$ spin symmetry which is not manifest in Abelian bosonization. In any event we will also
be interested in situations in which the $SU(2)$ symmetry is explicitly broken (say by the Zeeman
term) and in that case Abelian bosonization is the most direct way to solve this problem. 
Hence we proceed to use the standard Abelian bosonization approach in a similar manner
as in section \ref{sec:model} except that now the chiral Fermi fields 
are spin dependent. 

The right and left moving chiral Fermi fields with
spin $\alpha=\uparrow , \downarrow$ are bosonized according to the Mandelstam formulas
\begin{equation}
\psi^\dagger_{\pm\alpha}=\frac{1}{\sqrt{2\pi}}e^{\pm
i\phi_{\pm, \alpha}(x)} 
\label{eq:mandelstam-spin}
\end{equation}
where $\phi_{\pm, \alpha}$ are spin dependent chiral right and left moving
bose fields respectively. The corresponding bosonized normal-ordered density operators are 
\begin{equation}
J_{\pm , \alpha}=-{\D{\frac{1}{2\pi}}} \partial_x\phi_{\pm , \alpha}.
\label{eq:bosonization-spin}
\end{equation}
Extending the expression in Eq.~(\ref{eq:Lpm}) to the partially spin
polarized case of concern, the Lagrangians for the each spin component of the
decoupled non-interacting edges are 
\begin{equation}
{\mathcal L}_{\pm , \alpha}[\phi_{\pm , \alpha}]=\frac{1}{4\pi} 
\partial_x \phi_{\pm , \alpha} (\pm  \partial_t -v_0 \partial_x) \phi_{\pm , \alpha}. 
\label{eq:LpmS}
\end{equation}
The chiral boson fields $\phi_{\pm , \alpha}$ can be decomposed
into their spin and charge components:
\begin{equation}
\phi_{\pm, c}=\frac{1}{\sqrt{2}}(\phi_{\pm,\uparrow}+\phi_{\pm,\downarrow})
\quad
\phi_{\pm, s}=\frac{1}{\sqrt{2}}(\phi_{\pm,\uparrow}-\phi_{\pm,\downarrow})
\end{equation}
In terms of these chiral charge and spin bosons, the right moving electron operators are 
(up to Klein factors)
\begin{equation}
\psi_{+,\uparrow/\downarrow}^\dagger\sim {\D{\frac{1}{\sqrt{2\pi}}}} 
e^{\D{\frac{i}{\sqrt{2}} \phi_{c,+}}}
{e^{\D{\pm \frac{i}{\sqrt{2}} \phi_{s,+}}}}
\label{eq:electron-spin}
\end{equation}
\textit{i.\ e.\/} the electron splits into a spin-$1/2$ charge neutral spinon and a charge $1$ spin 0
holon.

The chiral charge currents $J_{c,\pm}$ are
\begin{equation}
J_{c,\pm}=-\frac{\sqrt{2}}{2\pi} \partial_x \phi_{c,\pm}
\label{eq:charge-current}
\end{equation}
The coefficient $\sqrt{2}$ in front of the charge current shows that the filling factor is
$\nu=2$. In what follows, exactly as what we found for fully polarized states,
 changes in the filling factor will only appear through the
dependence on $\nu$ of the coupling constants. However, the coefficient of the current will
remain unchanged. 
 
The corresponding expressions for the chiral spin currents $J_{a,\pm}$, $a=x,y,z$, 
the three generators of the 
$su(2)_1$ Kac-Moody algebra of spin, are
\begin{eqnarray}
J_{x,\pm}&=&\frac{1}{2\pi} \cos(\sqrt{2} \phi_{s,\pm})
\nonumber \\
J_{y,\pm}&=&\pm\frac{1}{2\pi} \sin(\sqrt{2} \phi_{s,\pm})
\nonumber \\
J_{z,\pm}&=&-\frac{1}{2\pi} \frac{1}{\sqrt{2}} \partial_x \phi_{s,\pm}
\nonumber \\
&&
\label{eq:spin-currents}
\end{eqnarray}
The factors of $\sqrt{2}$ are crucial for the system to be invariant under the $SU(2)$ symmetry 
of spin ~\cite{yellow}. 

In the absence of electron 
tunneling at the point contact, the Hamiltonian for the line junction reduces to 
$\mathcal H=\mathcal H_c+\mathcal H_s$ 
of Eq.\  (\ref{eq:Hc}) and Eq.\ (\ref{eq:Hs}) respectively. Thus
we recover the familiar  
spin-charge separation of one-dimensional 
interacting electronic systems. This Hamiltonian has been studied extensively in the literature 
(see for instance a pedagogical discussion 
in ref.\ ~\cite{spin}). The charge sector $\mathcal H_c$ behaves exactly as in the spin-polarized 
case of section \ref{sec:model}. The only difference here is the factor of $\sqrt{2}$ in the 
definition of the (bosonized) chiral charge currents which reflect the fact that these are 
the edge states of two quantum Hall states each with filling factor $\nu=2$. 
Thus the discussion of section \ref{sec:model} implies that the charge
sector is described by a rescaled charge boson $\varphi_c=(\phi_{c,+}+\phi_{c,-})/\sqrt{K_c}$, with Lagrangian
\begin{equation}
\mathcal L_c=\frac{1}{8\pi} \left(\frac{1}{v_c} \left(\partial_t \varphi_c\right)^2-v_c \left(\partial_x \varphi_c\right)^2\right)
\label{eq:Lc}
\end{equation}
with a charge Luttinger parameter $K_c$ equal to
\begin{equation}
K_c=\sqrt{\D {\frac{1-g_c}{1+g_c}}}.
\label{eq:Kc}
\end{equation}
Note that $K_c<1$ since $g_c>0$.
The compactification radius of the charge boson $\varphi_c$ is $R_c=\sqrt{2/K_c}$.
The velocity of the charge boson is renormalized exactly 
as in the spin-polarized case, \textit{ i.\ e.\/} $v_c=v_0 \sqrt{1-g_c^2}$. 

Naturally, the main difference between the case with a small spin
polarization and the fully polarized case 
resides in the spin sector with effective Hamiltonian $\mathcal H_s$. 
The first two terms of the $SU(2)$ symmetric part of spin Hamiltonian of Eq.\ (\ref{eq:Hs-symm}) represent 
two decoupled edges with exact $SU(2)$ symmetry. In fact, this is a fixed point Hamiltonian 
of two chiral $su(2)_1$ Wess-Zumino-Witten conformal field theories. Except for the 
renormalization of the velocities, due to forward scattering intra-edge interactions,  
this is a free theory. In Abelian bosonization the first two terms of the Hamiltonian of Eq.\
(\ref{eq:Hs-symm}), which we will denote by $\mathcal H_{s,\pm}$ are given by~\cite{yellow}
\begin{equation}
\mathcal H_{s,\pm}=\frac{2\pi}{3} v_s \vec J_{s,\pm}^2=\frac{v_s}{4\pi} (\partial_x \phi_{s,\pm})^2
\label{eq:Hspm}
\end{equation} 
The inter-edge exchange interaction term, with coupling constant 
$g_s$, is a chirality breaking perturbation and its effects are well
known~\cite{spin}. 
After Abelian bosonization, the Ising component exchange
interaction only renormalizes the velocity and the compactification
radius of the spin boson but the $XY$ component
exchange introduces cosine term as can be seen in the following
bosonized effective Lagrangian
\begin{eqnarray}
\mathcal L_s&=&\frac{1}{8\pi} \left(\frac{1}{v_s'} (\partial _t
\varphi_s)^2- v_s' (\partial_x \varphi_s)^2\right)\\ \nonumber
&&-\frac{v_s g_\perp}{\pi} \cos\left(\sqrt{2K_s}\varphi_s\right)
-\frac{\mu_BgB}{\pi}\sqrt{\frac{K_s}{2}}\partial_x\varphi_s,
\label{eq:Lvarphis}
\end{eqnarray}
where $g_\perp$ and $g_\parallel$ is defined in
Eq.~(\ref{eq:g-perp-g-par}) and the last term is the Zeeman term.
In Eq.~(\ref{eq:Lvarphis}), $\varphi_s$ is the rescaled spin boson
\begin{equation}
\varphi_s=(\phi_{s,+}+\phi_{s,-})/\sqrt{K_s}
\label{eq:varphpi-s}
\end{equation}
with the Luttinger parameter $K_s$, the renormalized velocity
$v_s'$ and the compactification radius of the spin boson given by
\begin{equation}
K_s=\sqrt{\frac{1-g_\parallel}{1+g_\parallel}},\quad
v_s'=v_s\sqrt{1-g_\parallel^2} \quad
R_s=\sqrt{2/K_s}.
\label{eq:Ks}
\end{equation}

\subsection{Small Zeeman term}
\label{sec:small-Zeeman}

\begin{figure}
\psfrag{xlabel}{$g_\parallel$}\psfrag{ylabel}{$g_\perp$}
\epsfig{file=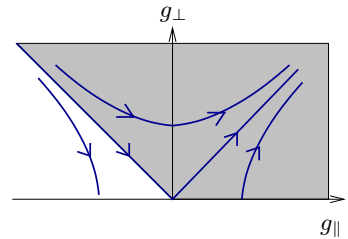,width=.25\textwidth}
\caption{The RG flow of Eq.~(\ref{eq:XY}). The trajectories starting
at points in the shaded region flow to spin gap phase.}
\label{fig:XY}
\end{figure}

Let us discuss first the case when the Zeeman energy is very small. 
Although 
this case does not apply to the samples used in the experiments of ref.\ \ref{ref:kang}, 
in which the Zeeman 
interaction is not small, nevertheless it is a good starting point for a theoretical analysis 
of this problem. 
The renormalization group (RG) $\beta$-functions for the exchange interaction
coupling constants are well known to have the following form~\cite{affleck,amit,geology}
\begin{eqnarray}
\D\frac{dg_\perp}{d \ln a}&=&2g_\perp g_\parallel-5g_\perp^3+\ldots
\nonumber \\
\D\frac{dg_\parallel}{d \ln a}&=&2g_\perp^2+4g_\perp^2g_\parallel+\ldots
\nonumber \\
&&
\label{eq:XY}
\end{eqnarray}
where $a$ is a length scale. The resulting RG flow is sketched in
Fig.~\ref{fig:XY}. 
%%%%%%%%%%%%%%%%%%%%%%%%%%%%%%%%%%%%%%%%%%%%%%%%%%%%%%%%%%%%%%%%%%
%%%%%%%%%%%%%%%%%%%%%%%%%%%%%%%%%%%%%%%%%%%%%%%%%%%%%%%%%%%%%%%%
The consequence of the flow depends on the anisotropy of interaction
and the sign of the coupling as the following.

\subsubsection{The $SU(2)$ symmetric case}
\label{subsubsec:su2}
This model
describes two $\nu=2$ \textit{singlet} quantum Hall states coupled along a line
junction. In this case we can define a single coupling constant  
$g_s\equiv g_\perp=g_\parallel$.
In spite of the anisotropic look of the Eq.~(\ref{eq:Lvarphis}), the
relation between $K_s$ and $g_\perp$, which is same as $g_\parallel$
in this case, guarantees that the allowed RG flows are
$SU(2)$ invariant.

~From the Eq.~(\ref{eq:XY}), the RG $\beta$-function for $g_s$
is 
\begin{equation}
\beta(g_s)=\frac{dg_s}{d\ln a}=2g_s^2-2g_s^3+\ldots
\end{equation}

For $g_s<0$, {\it i.\ e.\/} \emph{ferromagnetic} exchange coupling, the
cosine term is a {\sl marginally irrelevant} perturbation. Hence in the low
energy regime the effective coupling vanishes, $g_s \to 0$, 
albeit very slowly and giving rise to logarithmic corrections to scaling. 
Thus, for $g_s <0$  the spin sector of the line junction remains gapless 
and $K_s \to 1$, $R_s \to \sqrt{2}$, a result originally found by Luther and 
Emery~\cite{luther-emery} in the theory of the
one-dimensional electron gas. This is presumably the relevant case for the line junction in the
$SU(2)$-symmetric regime since the inter-edge exchange interaction is naturally ferromagnetic. 
However,
we will see below that magnetic anisotropy can make the antiferromagnetic regime accessible. 

In contrast, for $g_s>0$, {\it i.\ e.\/} 
\emph{anti-ferromagnetic} exchange coupling, this perturbation is
 \emph{marginally relevant} and 
the flow is asymptotically free. In this case the effective coupling constant $g_s$ 
flows to large values where scale invariance is violated. Hence, in this case the system 
flows to a phase with an energy gap in the spin sector, a \textit{spin
gap state}, along the $SU(2)$-invariant RG trajectory. 
This state is physically equivalent to the Haldane phase~\cite{haldane} of 
one-dimensional quantum Heisenberg antiferromagnets and to the Luther-Emery liquid 
of the one-dimensional electron gas~\cite{luther-emery}. 
In particular for $K_s=1/2$, the spin boson
$\varphi_s$ is equivalent to a massive fermion. This is the well known Luther-Emery point.

For small values of the coupling constant
$g_s$ the magnitude of the spin gap $\Delta_s$ can be determined by perturbative renormalization 
group methods. For a strictly $SU(2)$-invariant system the spin gap is the well known result
\begin{equation}
\Delta_s(g_s)=D \; \sqrt{g_s} \; e^{\displaystyle{-\frac{1}{2g_s}}}
\label{eq:Delta}
\end{equation}
where $D$ is an ultraviolet cutoff of the order of a fraction of the Fermi energy. (The factor of
$\sqrt{g_s}$ is due to corrections-to-scaling which appear at two loop order in $g_s$.) Given the
apparent smallness of the exchange coupling constant $g_s$, this result is probably good enough
here. For larger values of $g_s$ the spin gap can be determined either from the full Bethe-Anstaz
solutions of the sine-Gordon and Chiral Gross-neveu models, or at special points, such as the
Luther-Emery point, from bosonization arguments. In both cases in addition to the 
spin gap one finds a spectrum of solitons which should lead to interesting resonance
effects in tunneling.

\subsubsection{Effects of a Small Magnetic anisotropy}

Let us now discuss what happens
if there is a small magnetic anisotropy, \textit{i.\ e.\/} a small anisotropy in the exchange
interaction. Presumably for the samples used in the experiments of Kang and
coworkers~\cite{kang}, if there
is any anisotropy at all it is exceedingly small. However, we will
discuss this case here since it leads to interesting effects.
Magnetic anisotropy makes Ising exchange coupling $g_\parallel$ to
differ from $XY$-exchange $g_\perp$. In this case the RG flow no longer follows
$SU(2)$ invariant trajectory.
It is easy to see from the $beta$-functions Eq.~(\ref{eq:XY}) that for
$g_s>0$ (in which case both $g_\parallel$ and $g_\perp$ are positive),
the line junction will flow toward the spin gap state.

However, for $g_s<0$, the RG flows depend on the anisotropy.
With Ising like anisotropy  
($\lambda>0$) and $g_s<0$, the RG
trajectories flow toward the line of fixed points at zero
sine-Gordon coupling constant,  and $K_s >1$. 
Conversely with $XY$-like anisotropy ($\lambda<0$) we get the opposite result.
In this case, the RG trajectories still
flow initially toward the free theory ($g_\perp \to 0$). However,
they will eventually be driven to the marginally relevant flow of the $SU(2)$ trajectory
leaving the $su(2)_1$ fixed point. Hence, in this regime
the line junction flows toward the spin gap state. Thus, even though
the initial value of the inter-edge
interaction is negative, $g_s<0$, an arbitrarily small $XY$ anisotropy
drives the line junction necessarily to a spin gap state! This is a
remarkable effect which leads us to conclude that there is a phase
transition at $\lambda=0$.
The discussion above is summarized in the Fig.~\ref{fig:XY} where
the region in the coupling constant space which flows to the spin-gap
phase is shaded.

\subsection{The effect of Zeeman interactions}
\label{sec:zeeman}

Let us finally discuss the case of large Zeeman
interactions. Physically this is the most important case. It is also
the simplest. The charge sector is not affected by the Zeeman
interaction and it behaves exactly in the same way as in the previous cases. 
The effect of the Zeeman term on the spin sector depends on which
regime the line junction is in.  In the absence of a spin gap, which
as we saw above happens for $g_s<0$ in  the $SU(2)$ symmetric case or
with Ising-like magnetic anisotropy,  the cosine term is
irrelevant. In this case, the  
Zeeman term can be eliminated from the Lagrangian density by a shift
of the spin field:   
$\varphi_{s}\to\varphi_{s}+2\pi\gamma x/v_s$, where 
$\gamma \equiv\frac{\mu_BgB}{\pi}\sqrt{\frac{K_s}{2}}$. While this
shift has no effect  
on the charge sector it forces a twist in the boundary conditions of
the spin sector : 
\begin{equation}
\Delta\varphi_{s}\equiv\int dx\partial_x\varphi_{s}\to\Delta\varphi_{s}+
\frac{2\pi\gamma}{v_s}L
\label{eq:twist}
\end{equation}
where $ L$ is the length of the system (the barrier). Since $\partial_x{\phi}_s$ is 
proportional to spin density, the twist of BC's Eq.\ (\ref{eq:twist}) implies that the 
$z$-component of the spin polarization $M_z=\langle S_z \rangle$ is finite, 
$M_s\propto\gamma /v_s \propto B L$, and this state has a non-zero
spin polarization, although in general is not fully polarized. 

Hence, the only observable effect of the Zeeman  term in gapless phase
is a
non-zero spin polarization and hence a twist of  the boundary
conditions. 

Let us now discuss the effects of the Zeeman term in the spin gap
phase. An examination of the effective Lagrangian $\mathcal{L}_s$ of
Eq.~(\ref{eq:Lvarphis}) shows that, as expected, there is a competition
between the Zeeman term and the cosine operator. This competition,
which bears a close analogy with the mechanism of the
commensurate-incommensurate transition, leads to different physical
behaviors depending on which is the smallest energy scale, the Zeeman
energy or the spin gap. When the Zeeman energy is small compared with
the spin gap, the system will stay in gapped phase despite the twist
of BC's. However, when the Zeeman term
dominates, the cosine operator once again become irrelevant and the
spin gap is destroyed by the Zeeman interaction. 

\subsection{Tunneling transport}

The discussions in the previous two subsections can be summarized as the
following. Depending on the sign of the exchange interactions,
magnetic anisotropy and the strength of Zeeman term, the spin sector
of the system can be
either in a spin gap phase or a gapless phase.
Let us finally look at the consequences of these results for the
question of  electron tunneling transport  
in the line junction. 
Due to the spin degree of freedom there are now three possible types
of tunneling corresponding to tunneling of charge and/or spin degree
of freedom. The lowest order operators for each of these processes are the
single electron tunneling operator which transports both charge and spin:
\begin{eqnarray}
{\mathcal O_e}&=&\psi_{\uparrow, +}^\dagger \psi_{\uparrow, -}+
\psi_{\downarrow, +}^\dagger\psi_{\downarrow, -}+{\rm h.\; c.\; }\nonumber\\
&\propto& \cos\left(\sqrt{\frac{K_c}{2}}\varphi_c\right)  
\cos\left(\sqrt{\frac{K_s}{2}}\varphi_s\right)\ ,
\label{eq:etunnel}
\end{eqnarray}
the spin singlet pair (spin 0, charge2) tunneling operator:
\begin{eqnarray}
{\mathcal O_{pair}}&=&\psi_{\uparrow,+}^\dagger\psi_{\downarrow,+}^\dagger
\psi_{\downarrow,-}\psi_{\uparrow,-}+{\rm h.\ c.\ }\nonumber\\
&\propto&\cos\left(\sqrt{2K_c} \; \varphi_c
\right)\ ,
\end{eqnarray}
and the tunneling operator of a spin 1 charge neutral excitation:
\begin{eqnarray}
{\mathcal O_s}&=&\psi_{\downarrow,-}^\dagger \psi_{\downarrow,+}
\psi_{\uparrow,+}^\dagger \psi_{\uparrow,-}+{\rm h.\ c.\ }\nonumber\\
&\propto&\cos\left(\sqrt{2K_s}\; \varphi_s\right).
\end{eqnarray} 
The single electron tunneling operator $\mathcal O_e$
clearly mixes the charge and spin sectors. As far as the charge sector is concerned, this
tunneling operator is similar to the one for fully polarized electrons shown in Eq.\ (\ref{Lfinal}),
except that instead of the Luttinger parameter $K$ we now have $K_c$, where $K_c$
is the charge Luttinger parameter defined in Eq.\ (\ref{eq:Kc}). The spin sector has a similar
structure with the effective Luttinger parameter $K_s$. The
scaling dimension of the operator of Eq.\ (\ref{eq:etunnel}) at a
point contact is
\begin{equation}
d_e={\D{\frac{1}{2}(K_c+K_s)}}.
\label{eq:e-dimension}
\end{equation}
The singlet pair tunneling operator ${\mathcal O_{pair}}$ which depends only
on the charge boson  and the holon pair tunneling operator ${\mathcal
O_s}$ which depends only on the spin boson are higher order operators. At
a point contact,  ${\mathcal O_{pair}}$ and ${\mathcal
O_s}$ have boundary scaling dimension $d_{pair}$ and $d_s$
respectively given by 
\begin{equation}
d_{pair}=2K_c\ ,\qquad d_s=2K_s.
\label{eq:pairs-dimension}
\end{equation}

Now let us discuss the possible effect of these operators in the spin-gap
phase and the gapless phase. First, because the singlet pair tunneling operator ${\mathcal O_{pair}}$
depends only on the charge boson, its effect is the same for the
spin-gap phase and the gapless phase. Since the charge sector is free,
the constraint of momentum conservation forbids the singlet pair
tunneling in the absence of a point contact. However, the
operator ${\mathcal O_{pair}}$  at a point contact is relevant 
for $K_c<1/2$ in the presence of strong Coulomb interaction (Eq.~(\ref{eq:pairs-dimension}))
and it can
lead to  charge only tunneling for both spin-gap phase and gapless
phase.
On the other hand, the possibilities of the other two
tunneling processes, namely the single electron tunneling and the
holon pair tunneling, depend on the presence or absence of the spin gap
since their operator representation involves vertex operators of spin
bosons.

\subsubsection{The spin-gap phase}

In the gapped phase, the spin boson field  $\varphi_s$ acquires an
expectation value in the set $\varphi_s=2n\pi /\sqrt{2K_s}$
where $n \in \mathbb Z$ 
which labels the manifold of degenerate
ground states in the gapped phase.  
Since the value of  $\cos({\sqrt{K_s/2}}\; \varphi_s)$ alternates in this
set, the expectation value of $\mathcal O_e$ vanishes in this phase and
the single electron tunneling is (exponentially) suppressed in
this regime. (This is a natural result since the electron 
carries spin $1/2$.) Therefore, the lowest order tunneling process
that can contribute to a charge transport across the barrier is the
singlet-pair tunneling, which is possible only through a point contact
for both spin-gap phase and gapless phase. Although this is a two
particle process, $\mathcal O_{pair}$ can still lead to a ZBC peak even
in this spin gap phase if
Coulomb interaction is strong enough so that $K_c<1/2$ which makes this operator relevant as
we mentioned earlier.
However since ${\mathcal O_s}$ is relevant and allowed everywhere
along the barrier in the spin gap phase ($\mathcal O_s$ is the
operator that causes the spin gap), there is a perfect spin tunneling in
the absence of charge tunneling even in the absence of a
point contact.
The mechanism behind this effect in the spin sector is
similar in spirit 
to the explanation of the ZBC peak in the \textit{charge}
tunneling  conductance proposed by Mitra 
and Girvin~\cite{girvin}. 
In fact this phase looks very much like a superconductor without phase 
coherence~\cite{stripe}.

\subsubsection{The spin gapless phase}

\begin{figure}[t!]

\epsfig{file=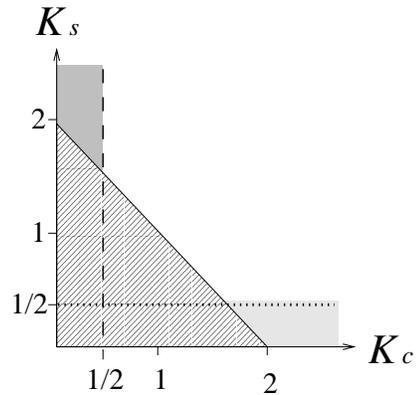, width=.3\textwidth}
\caption{The phase diagram near $\nu\sim2$.
      The peak in both spin tunneling conductance and the charge 
      tunneling conductance is expected in the cross-hatched region 
      below the line $K_c\!+\!K_s=2$ where the single particle
      tunneling operator is relevant.
      On the other hand, the charge only tunneling operator ${\mathcal O}_c$
      and the spin only tunneling operator ${\mathcal O}_s$ respectively
      are relevant in the dark shaded region to the left of the dashed line,
      lightly shaded region below dotted line.}\label{fig:spinful-phase}
\end{figure}

In the gapless phase, ${\mathcal O_e}$ or ${\mathcal O_s}$ also are
allowed only at a point contact and whether any of these operators are
relevant or not depends on the Luttinger parameters.
The effect of three tunneling operators ${\mathcal O_e}$, ${
\mathcal O_{pair}}$ and ${\mathcal O_s}$ at a point contact in the gapless phase is 
summarized in the Fig.~\ref{fig:spinful-phase} as pointed  out earlier
by Kane and Fisher~\cite{kane}. The cross-hatched
region is where the single 
particle tunneling operator ${\mathcal O_e}$ is relevant and we expect the peak in both
spin tunneling conductance and the charge tunneling conductance. In
the dark shaded region to the left of the dashed line, the charge only 
tunneling operator ${\mathcal O}_{pair}$ is relevant. Analogously, the spin only
tunneling operator ${\mathcal O}_s$ is relevant in the lightly shaded region
below the dotted line. Note that one has to keep in mind
that $K_c<1$ because of Coulomb interaction. 

For the $SU(2)$ 
symmetric gapless case (with ferromagnetic exchange) in which $K_s=1$,
the (boundary) scaling dimension of the electron tunneling  operator is 
$(K_c+1)/2<1$, since $K_c<1$. Thus, the single electron tunneling term is a relevant
perturbation, and the coupling constant $\Gamma$ flows to 
strong coupling in this case. Therefore, there should also be a zero-bias peak in the tunneling
conductance in the case of a $\nu=2$ 
spin singlet quantum Hall state, with qualitatively similar properties
as the zero bias tunneling peak for the spin polarized case.
With
ferromagnetic exchange interactions and Ising like anisotropy, in which case the system is in
gapless phase independent of the strength of Zeeman term, $K_s>1$ and
Fig.~\ref{fig:spinful-phase} implies that holon pair tunneling is
always irrelevant in this case. However, if $1<K_s<2-K_c$ for weak
ferromagnetic interaction, the single
electron tunneling is relevant. Also with strong enough Coulomb
interaction, singlet pair tunneling can become relevant.
Finally, for a phase in which the gap is washed out due to a strong
Zeeman term, $K_s<1$ and  again the single electron tunneling is relevant.

%%%%%%%%%%%%%%%%%%%%%%%%%%%%%%%%%%%%%%%%%%%%%%%%%%%%%%%%%%%%%%%%%%%%%%%%%%%%%%%%%%%%%%%%%%%
%%%%%%%%%%%%%%%%%%%%%%%%%%%%%%%%%%%%%%%%%%%%%%%%%%%%%%%%%%%%%%%%%%%%%%%%%%%%%%%%%%%%%%%%%%%

To summarize, 
in both spin-gap phase and gapless phase,
there is no charge tunneling current in the absence of a point
contact. In the gapless phase, there underlies a quantum phase
transition in the single electron
tunneling process through a
point contact which leads to the reappearance of the zero
bias peak near $\nu\sim2$ in a manner similar to the fully polarized
case of the previous section. On the other hand, even though the
single electron tunneling is exponentially suppressed in the spin-gap
phase, an analogous crossover in the singlet pair tunneling channel 
can lead to the 
reappearance of the ZBC peak in the (in the regime of strong
Coulomb interactions).
This scenario is consistent with the experimental observation which
displays a very close similarity between the manner in which peak region
appears abruptly and disappears gradually in both $\nu\sim1$ and $\nu\sim2$.  
Since the operator ${\mathcal O_e}$ mixes the spin and
the charge sectors, in the regime in which this operator is relevant it induces 
a non-zero tunneling current of both charge and spin.  
Hence if the observed ZBC peak near $\nu\sim2$ is
indeed caused by the single 
particle tunneling operator in the phase without a spin gap,
we expect that a \textit{spin conductance} peak should be observable near $\nu\!\sim\!2$
but not near $\nu\!\sim\!1$, in marked contrast to charge conductance which 
would show a ZBC peak near both filling factors.

In this section, we extended the picture we advocated for in the
previous sections to the case of small spin polarization near $\nu\sim2$ and
investigated the changes in physics brought about by the electron
spin.
It was pointed out that the interplay between the Zeeman term and  the
exchange term enables us to identify two 
different phases in terms of their
spin transport properties even in the absence of any point contact operator:
a spin-gap phase in which the spin excitations along the edge are
gapped, and hence perfect spin tunneling, and a phase with
gapless spin excitations. In both cases,
there is no charge tunneling current in the absence of a point
contact. Since in most cases of physical interest the edge states are likely to be in the
gapless phase at least for a large enough Zeeman interaction, we
proposed that here too there is a crossover in single electron tunneling processes through a
point contact leading to the reappearance of the zero
bias peak near $\nu\sim2$. In our picture, the apparent similarities
in the patterns in which the peak region begins and disappears near
two filling factors $\nu\sim1$ and $\nu\sim2$ in the experiment by
Kang and coworkers can be understood in a
natural and consistent way. The reappearance of
peak region near $\nu\sim2$ had  been totally unexplained in previous theories of tunneling 
between laterally coupled FQH states~\cite{girvin,sachdev}. We also discussed a number of 
interesting two-particle tunneling processes and the interesting behavior of spin tunneling 
in these systems.
%%%%%%%%%%%%%%%%%%%%%%%%%%%%%%%%%%%%%%%%%%%%%%%%%%%%%%%%%%%%%%%%%%%%%%%%%%%%%%%%%%%%%%%%%%%%%
\section{Conclusions}
\label{sec:conclusion}
%%%%%%%%%%%%%%%%%%%%%%%%%%%%%%%%%%%%%%%%%%%%%%%%%%%%%%%%%%%%%%%%%%%%%%%%%%%%%%%%%%%%%%%%%%%%%

In summary, in this paper we proposed a theoretical explanation of the questions
raised by experiments of Kang and coworkers~\cite{kang}, by modeling the system
as a pair of coupled chiral Luttinger liquid with a point contact.
Using standard bosonization methods we mapped the  problem  to the
tunneling problem in Luttinger liquids first discussed by Kane and Fisher 
~\cite{kane,extra-kane}. Our results show that the inter-edge Coulomb
interaction reduces the Luttinger 
parameter and moves the system deep into the strong coupling
regime for $\nu\sim1$ leading to the appearance of zero-bias peak in
the tunneling conductance with a peak value at $T=0$ of
$G_t=Ke^2/h$, with $K<1$. 
We mapped the problem to integrable boundary sine-Gordon theory, and
used the known exact results of BSG problem to obtain predictions for
the behavior of the tunneling conductance. By considering a special
solvable case, we determined the behavior of the conductance for all
temperature and voltages. We investigated several crossovers of
interest by introducing an appropriate $\beta$-function. This analysis
showed that the crossover between the $T=0$ behavior and the $V=0$
behavior yields a natural explanation of the low value of the ``zero
bias conductance peak'' measured in the experiment~\cite{kang}. 
We also showed that the gradual disappearance of the peak 
as the filling factor is increased past $\nu\!\sim\!1$ can
be ascribed to the crossover between $T\!<\!T_K$ and $T\!>\!T_K$.

Furthermore, we considered the role of spin in this tunnel junctions and showed that
the reappearance of the ZBC peak in the region near filling factor $\nu\sim2$ can be understood 
if we assume that there is a (possibly small) spin polarization near
$\nu\sim2$. We extended
the approach we used for fully polarized electrons with $\nu\sim1$ to
partially spin polarized and unpolarized electrons with $\nu\sim2$, by
taking into account the role of Zeeman interactions, exchange
interactions and magnetic anisotropies. 
We discussed in detail the phase diagram of the system in this case
and showed that the tunneling signature depends on whether the spin
sector is gapped or not. We showed that the picture near $\nu\sim1$
can be naturally extended to this new regime, and that the single
particle tunneling operator can also give rise to a zero-bias tunneling
conductance peak in both charge transport and spin transport in the
gapless phase. 
Higher order (multi-particle) point contact operators  can
in principle lead to charge only or spin only tunneling, depending on the value of
Luttinger parameters $K_c$ and $K_s$. %Howveer spin only tunneling is
%unrealistic in the specific case relevant to the experiment by Kang and coworkers.  
On the other hand, we found that spin transport along the edge is
gapped even in the absence of point contact when Zeeman term is
small and if there is a very weak $XY$-like magnetic anisotropy or if
the exchange interaction is anti-ferromagnetic.
In this regime we expect perfect tunneling of the spin
current  
which suggest future experimental tests of these ideas. Even though
the single electron tunneling is exponentially suppressed in the
spin-gap phase, the singlet pair tunneling can lead to a ZBC peak in
the presence of strong Coulomb interaction.

Our scenario is based on the assumption of 
single tunneling center. When the bias voltage and and the coupling between 
edge modes on each side of the barrier are weak enough to give a
low peak in the tunneling conductance, as is observed in the experiment,
the scenario of tunneling through a single tunneling center
is quite likely to be an accurate description of the physics. 
Even though our picture is applicable only near (and above) $\nu=1$
and $\nu=2$, it offers a 
natural explanation of many salient features of the experiment which
were not explained 
so far. This picture offers a consistent explanation for the reappearance
of the ZBC peak and of the observed similarity in the manner in which
the two peak 
regions near $\nu\sim1$ and $\nu\sim2$ appear and disappear.  Our
results also indicate that temperature should play an important role ,
and that a temperature dependence of the data is needed to understand
what is going on. In particular we predict that  
 as temperature is lowered the crossover filling factor
$\nu^*$ will be lowered, and that the width of the peak region (in
filling factor) 
as well as the height of the the ZBC peak will
increase. 
We also anticipate a region with
ZBC peak in spin conductance near $\nu\!\sim\!2$ but not near $\nu\!\sim\!1$.
We find that there are more than one mechanisms through which spin tunneling can
happen and depending on the channel, the spin tunneling may or may not
be accompanied by charge tunneling.

Although in this paper we considered only the simplest possible case of a single tunneling center
it is interesting to investigate the effects of more than one impurity. While we have not
investigated this problem extensively, it is clear that there should be interesting interference effects if
there is more than one tunneling center. Indeed, some time ago Chamon and
coworkers~\cite{interferometer} proposed an experiment based on a two-tunneling center device 
in the fractional quantum Hall regime as a way to measure the fractional statistics of Laughlin
quasiparticles.

We finally note that there is a  recent paper by Carpentier, Peca and
Balents~\cite{carpentier} on a related problem.
Carpentier and coworkers calculated the tunneling current between
interacting Luttinger liquids constructed in a similar geometry as the
geometry of experiment by Kang and coworkers. They showed that electron
fractionalization can be probed from multiple branch points of current
density. However, both the effect of charging(leaking) from(to) the bulk
system and the absence of chirality constraint make the system
considered in the Ref.~\cite{carpentier} quite different from the
system considered in this paper in connection to the experiment by
Kang and coworkers.

\begin{acknowledgments}
We thank Prof.\ W.\ Kang for several useful and stimulating discussions.  
This work was supported in part by the National
Science Foundation through the grants No.\ DMR98-17941 and DMR01-32990.
\end{acknowledgments}

%%%%%%%%%%%%%%%%%%%%%%%%%%%%%%%%%%%%%%%%%%%%%%%%%%%%%%%%%%%%


\begin{thebibliography}{99}
%%%%%%%%%%%%%%%%%%%%%%%%%%%%%%%%%%%%%%%%%%%%%%%%%%%%%%%%%%%%

\bibitem{hal}
B.\ I.\ Halperin, Phys.\ Rev.\ B {\bf 25}, 2185 (1982).
\label{ref:hal}

\bibitem{wen}
X.\ G.\ Wen, Phys.\ Rev.\ B {\bf 41}, 12838 (1990).
\label{ref:wen}

\bibitem{stone}
M.\ Stone, (1990).
\label{ref:stone}

\bibitem{kane}
C.\ L.\ Kane and M.\ P.\ A.\ Fisher, Phys.\ Rev.\ B {\bf 46}, 15233 (1992).
\label{ref:kane}

\bibitem{extra-kane}
C.\ L.\ Kane and M.\ P.\ A.\ Fisher,
Phys.\ Rev.\ Lett. {\bf 68}, 1220(1992); {\it ibid}, {\bf 72}, 724 (1994).
\label{ref:extra-kane}

\bibitem{milliken}
F.\ P.\ Milliken, C.\ P.\ Umbach, and R.\ A.\ Webb, 
Solid State Comm.\ {\bf  97}, 309 (1996).
\label{ref:milliken}

\bibitem{chang}
A.\ M.\ Chang, L.\ N.\ Pfeiffer, and K.\ W.\ West, 
Phys.\ Rev.\ Lett.\ {\bf 77}, 2538 (1996); 
M. Grayson, D.\ C.\ Tsui, L.\ N.\ Pfeiffer, K.\ W.\ West, and A.\ M.\ Chang, 
{\it  ibid}, {\bf 80}, 1062 (1998); 
A.\ M.\ Chang, M.\ K.\ Wu, C.\ C.\ Chi, L.\ N.\ Pfeiffer, and K.\ W.\ West, 
{\it ibid}, {\bf 86}, 143 (2001).
\label{ref:chang}

\bibitem{chamon}
C.\ de C.\ Chamon and E.\ Fradkin, Phys.\  Rev.\ B {\bf 56}, 2012 (1997).
\label{ref:chamon}

\bibitem{shytov}
A.\ V.\ Shytov, L.\ S.\ Levitov and B.\ I.\ Halperin, 
Phys.\ Rev.\ Lett.\ , {\bf 80}, 141 (1998). 
\label{ref:shytov}

\bibitem{ana}
Ana L\'opez and Eduardo Fradkin, Phys.\  Rev.\ B {\bf 59}, 15323 (1999).
\label{ref:ana}

\bibitem{joel}
J.\ Moore and X-G.\ Wen, Phys.\ Rev.\ B {\bf 57}, 10138 (1998). 
\label{ref:joel}

\bibitem{kang}
W.\ Kang, H.\ L.\ Stormer, L.\ N.\ Pfeiffer, K.\ M.\ Baldwin, and K.\ W.\ West, 
Nature {\bf 403}, 59 (2000).
\label{ref:kang}

\bibitem{ho}
T.\ -L.\ Ho, Phys. Rev. B,{\bf 50}, 4524 (1994).
\label{ref:ho}


\bibitem{girvin}
A.\ Mitra and S.\ M.\ Girvin, Phys.\ Rev.\ B {\bf 64}, R41309 (2001).
\label{ref:girvin}

\bibitem{lee-yang}
H.\ C.\ Lee and S.\ R.\ Eric Yang, Phys.\ Rev.\ B {\bf 63}, (BR) 199308-1 (2001).
\label{ref:lee-yang}

\bibitem{sachdev}
M.\ Kollar and S.\ Sachdev, Phys.\ Rev.\ B {\bf 65}, 121304(R) (2002).
\label{ref:sachdev}

\bibitem{takagaki}
Y.\ Takagaki and K.\ H.\ Ploog, Phys. Rev. B, {\bf 62}, 3766 (2000).
\label{ref:takagaki}

\bibitem{moon}
K.\ Moon, H.\ Yi, C.\ L.\ Kane, S.\ M.\ Girvin, and M.\ P.\ A.\ Fisher, 
Phys.\ Rev.\ Lett.\  {\bf 71}, 4381 (1993).
\label{ref:moon}

\bibitem{wen9091}
X.\ G.\ Wen, Phys. Rev. Lett. {\bf 64}, 2206 (1990); Phys. Rev. B {\bf 43}, 11025 (1991); 
{\it ibid}, {\bf 44}, 5708 (1991).
\label{ref:wen9091}


\bibitem{kang-priv}
W.\ Kang, private communication.
\label{ref:kang-priv}

\bibitem{bosonization}
See, for instance, {\sl Bosonization} by Michael Stone, World Scientific (Singapore, 1994), and
references therein; V.\ J.\ Emery , {\sl Theory of the One-Dimensional Electron Gas}, in{\sl Highly
Conducting One-Dimensional Solids}, Edited by J.\ T.\ Devreese \textit{et.\ al.\/}, Plenum Press, New
York, 1979); {\sl Bosonization and Strongly Correlated Systems} by A.\ O.\ Gogolin, A.\ A.\ Nersesyan
and A.\ M.\ Tsvelik, Cambridge University Press (Cambridge, 1998); 
{\sl An Introduction to Bosonization} by David S\'en\'echal, in the {\sl Proceedings of the Workshop
on Theoretical Methods for Strongly Correlated Fermions}, Centre de Recherches Math\'ematiques in
Montr\'eal, Canada (cond-mat/9908262).
\label{ref:bosonization}
%\bibitem{sadao}
%S.\ Adachi, J. Appl. Phys. {\bf 58}(3),R1 (1985).
%\label{ref:sadao}

\bibitem{fendley}
P.\ Fendley, A.\ W.\ W.\ Ludwig, and H. Saleur, Phys.\ Rev.\ B {\bf 52}, 8934 (1995); 
P.\ Fendley, A.\ W.\ W.\ Ludwig, and H.\ Saleur, Nucl.\ Phys.\  B {\bf 45A}, 29 (1996).
\label{ref:fendley}

\bibitem{saleur}
H.\ Saleur, {\sl Lectures on Non Perturbative Field Theory and Quantum Impurity Problems},
Proceedings of the 1998 Les Houches Summer School on {\sl Topological Aspects of
Low Dimensional Systems},  (Les Houches, Haute Savoie, France; 1998).
\label{ref:saleur}

\bibitem{affleck}
I.\ K.\ Affleck and A.\ W.\ W.\ Ludwig, Nucl.\ Phys.\ B {\bf 352}, 849 (1991).
\label{ref:affleck}

\bibitem{polchinski}
J.\ Polchinski and L.\ Thorlacius, Phys.\ Rev.\ D {\bf 50}, 622 (1994).
\label{ref:polchinski}

\bibitem{andrei}
N.\ Andrei, K.\ Furuya abd J.\ H.\ Lowenstein, Rev.\ Mod.\ Phys.\ {\bf 55}, 331 (1983).
and references therein.
\label{ref:andrei}


\bibitem{ghoshal}
S.\ Goshal and A.\ Zamolodchikov, Int.\ Jour.\ Mod.\ Phys.\ B {\bf 9}, 3841 (1994).
\label{ref:ghoshal}

\bibitem{ludwig-affleck}
I.\ K.\ Affleck and A.\ W.\ W.\ Ludwig, Phys.\ Rev.\ Lett.\ {\bf 67}, 3160 (1991); 
A.\ W.\ W.\ Ludwig
and  I.\ K.\ Affleck, Phys.\ Rev.\ Lett.\ {\bf 68}, 1046 (1992); 
Phys.\ Rev.\ B {\bf 48}, 7297 (1993)
\label{ref:ludwig-affleck}

\bibitem{fradkin}
Eduardo Fradkin, {\sl Exploring the fractional quantum Hall effect with 
electron tunneling}, in {\sl Quantum Physics at the Mesoscopic Scale},
Proceedings of the XXXIVth Rencontres de Moriond, 
Edited by C.\ Glattli, M.\ Sanquer and J.\ Tr{\^a}nh V{\^a}n,
Les Arcs, Haute Savoie, France , January 1999, EDP Sciences 
(Les Ulis, France, 2000);
cond-mat/9905218.
\label{ref:fradkin}


\bibitem{weiss}
U.\ Weiss, Solid State Comm.\  {\bf100}, 281 (1996).
\label{ref:weiss}

\bibitem{zwerger}
M.\ P.\ A.\ Fisher and W.\ Zwerger, Phys.\ Rev.\ B {\bf 32}, 6190 (1985)
\label{ref:zwereger}

%\bibitem{fradkin_FT}
%E.\ Fradkin,{\it Field Theories of Condensed Matter Systems}, Addison Wesley
%(Redwood City, 1991).
%\label{ref:fradkin_FT}

%\bibitem{fradkin_susskind}
%E.\ Fradkin and L.\ Susskind, Phys.\ Rev.\ D {\bf 17}, 2037 (1978).
%\label{ref:fradkin_susskind}

%\bibitem{polyakov}
%A.\ M.\ Polyakov, Phys.\ Lett.\ {\bf 59B}, 79 (1975).
%\label{ref:polyakov}


\bibitem{koutouza}
A.\ Koutouza, F.\ Siano and H.\ Saleur, 
J.\ Phys.\ A{\bf 34}, 5497 (2001).
\label{ref:koutouza}

\bibitem{subir}
See for instance, {\sl Quantum Phase Transitions} by Subir Sachdev, 
Cambridge University Press (1998).
\label{ref:subir}

%\bibitem{kang-new}
%W.\ Kang {\it et.\ al.\/}, unpublished. We are grateful to Professor Kang for making this
%unpublished data available to us.
%\label{ref:kang-new}



\bibitem{chamon_freed}
C.\ de C.\ Chamon, D.\ E.\ Freed, and X.\ G.\ Wen, Phys. Rev. B {\bf 51}, 2363 (1995); 
{\it ibid}, {\bf 53}, 4033 (1996).
\label{ref:chamon_freed}



%\bibitem{comment}
%Our curve give an excellent fit of the data up to tempereratures $T \sim 9 K \sim 2 T_K$, where
%the data shows a tendency to curve upwards. Although this effect becomes more pronounced at
%smaller values
%of $K$, it affects only the high temperature behavior of the data. At the present time this effect 
%cannot be explained by our theory.
%\label{ref:comment}

\bibitem{ana2}
See, for instance, 
Ana L\'opez and Eduardo Fradkin, Phys.\  Rev.\ B {\bf 63}, 085306 (2001), and references therein.
\label{ref:ana2}

\bibitem{spin}
This decomposition has been used often in the literature. See for instance the 
reviews {\sl Field Theory Methods and Quantum Critical Phenomena} by I.\ K.\ Affleck, in
\textit{Fields, Strings and Critical Phenomena}, edited by E.\ Br\'ezin and J.\ Zinn Justin,
Les Houches Summer School on Theoretical Physics, Session XLIX (North-Holland, Amsterdam, 1990). 
\label{ref:spin} 

\bibitem{awschalom}
G.\ Salis, D.\ D.\ Awschalom, Y.\ Ohno and H.\ Ohno, Phys.\ Rev.\ B {\bf 64}, 195304 (2001).
\label{ref:awschalom} 

\bibitem{yellow}
See, for instance, {\sl Conformal Field Theory} by P.\ Di Francesco, P.\ Mathieu and 
D.\ S\'en\'echal,
Springer-Verlag (New York, 1996); P.\ Ginsparg, {\sl Applied Conformal Field Theory}, 
in \textit{Fields, Strings and Critical Phenomena}, edited by E.\ Br\'ezin and J.\ Zinn Justin,
Les Houches Summer School on Theoretical Physics, Session XLIX (North-Holland, Amsterdam, 1990). 
\label{ref:yellow} 

\bibitem{amit}
D.\ J.\ Amit, Y.\ Y.\ Goldschmidt and G.\ Grinstein, J.\ Phys.\ A {\bf 13}, 585 (1980).
\label{ref:amit}

\bibitem{geology}
This result is alos well known from the g-ology literature. See, for instance, J.\ S\'olyom, Adv.\
in Phys.\ {\bf 28}, 201 (1979), and V.\ J.\ Emerey's review cited in Ref.\ \cite{bosonization}.
\label{ref:geology}


\bibitem{luther-emery}
A.\ Luther and V.\ J.\ Emery, Phys.\ Rev.\  Lett.\ {\bf 33}, 589 (1974).
\label{ref:luther-emery} 

\bibitem{haldane}
F.\ D.\ M.\ Haldane, Phys.\ Lett.\  A {\bf 93}, 464 (1983); 
Phys.\ Rev.\ Lett.\ {\bf 50}, 1153 (1983); J.\  Appl.\  Phys.\ {\bf 57}, 3359 (1985).
\label{ref:haldane}

\bibitem{stripe}
This argument has been made by V.\ J.\ Emery, S.\ A.\ Kivelson and O.\ Zachar, 
Phys.\ Rev.\ B{\bf 56}, 6120 (1997), in the context of the spin gap proximity effect in 
striped superconductors.
\label{ref:stripe}

%\bibitem{kosterlitz}
%J.\ Kosterlitz and D.\ Thouless, J. Phys. C: Solid State Phys. {\bf 6}, 1181 (1973)
%\label{ref:kosterlitz}

\bibitem{interferometer}
C.\ de C.\ Chamon, D.\ E.\ Freed, S.\ A.\ Kivelson, S.\ L.\ Sondhi, and X.\ G.\ Wen,
Phys.\ Rev.\ B {\bf 55}, 2331(1997).
\label{ref:interferometer}

\bibitem{carpentier}
D.\ Carpentier, C.\ Pe\c ca, and L.\ Balents, cond-mat/0103193.
\label{ref:carpentier}
\end{thebibliography}
\end{document}